\documentclass[twocolumn]{aastex701}

\usepackage{url} 
\usepackage{cancel}
\usepackage{hyperref}

\usepackage{graphicx}
\usepackage{subcaption}
\usepackage{enumitem}

\begin{document}

\title{The James Webb Space Telescope Absolute Flux Calibration. VI. Near-Infrared Camera Imaging and Coronagraphy}

\author[0000-0003-4850-9589]{Martha L.\ Boyer}
\affiliation{Space Telescope Science Institute, 3700 San Martin Drive, Baltimore, MD 21218, USA}
\email{mboyer@stsci.edu}

\author[0000-0003-3759-8707]{Benjamin Sunnquist}
\affiliation{Space Telescope Science Institute, 3700 San Martin Drive, Baltimore, MD 21218, USA}
\email{bsunnquist@stsci.edu}

\author[0000-0002-6875-1543]{Bryan Hilbert}
\affiliation{Space Telescope Science Institute, 3700 San Martin Drive, Baltimore, MD 21218, USA}
\email{hilbert@stsci.edu}

\author[0000-0001-7410-7669]{Dan Coe}
\affiliation{Space Telescope Science Institute, 3700 San Martin Drive, Baltimore, MD 21218, USA}
\email{dcoe@stsci.edu}

\author[0009-0008-4009-3391]{Varun Bajaj}
\affiliation{Space Telescope Science Institute, 3700 San Martin Drive, Baltimore, MD 21218, USA}
\email{vbajaj@stsci.edu}

\author[0000-0001-8354-7279]{Paul Bennet}
\affiliation{Space Telescope Science Institute, 3700 San Martin Drive, Baltimore, MD 21218, USA}
\email{pbennet@stsci.edu}

\author[0000-0001-8627-0404]{Julien Girard}
\affiliation{Space Telescope Science Institute, 3700 San Martin Drive, Baltimore, MD 21218, USA}
\email{jgirard@stsci.edu}


\begin{abstract}
  We present an updated flux calibration for all imaging modes of the Near-Infrared Camera on JWST that converts instrumental units to physical surface brightness units of MJy sr$^{-1}$. This calibration includes observations of 19 flux standard stars spanning 3.5 years, with a mix of A dwarfs, solar analogs, and hot stars. All 5 coronagraphic setups, all 29 filters, and both weak lenses are calibrated. This is the first on-sky calibration for the weak lenses used in Time Series observations and for secondary coronagraphic configurations (long wavelength masks paired with short wavelength filters, and vice versa). We also assess count-rate differences between subarrays and the full frame, finding differences of up to $\sim$1\%. These differences are incorporated into the calibration factors.  We find that the scatter in the calibration factor is typically $<$2\%, with about half of the filter+detector[$+$mask] combinations reaching $<$1\% scatter. There are no trends with detector effects such as the count rate and well depth.  Images in a handful of filters of the Large Magellanic Cloud and globular cluster 47 Tuc show that residual detector-to-detector offsets are typically small ($<$1\%), but can be as high as 4--5\%. The NIRCam detectors are found to be quite stable, with possible count-rate decreases of $<$0.4\% per year, which is within the calibration uncertainties. These new calibration factors were incorporated into the JWST pipeline in 2026 March.

\end{abstract}


\section{Introduction}

Absolute flux calibration is an essential part of an instrument's calibration that converts astronomical measurements to physical units.  \citet{Gordon+2022} describes the overall flux calibration plan for all instruments onboard the James Webb Space Telescope (JWST), which covers wavelengths spanning 0.6 -- 29~$\mu$m \citep{Gardner+2023, Rigby+2023}.    The absolute flux calibration program is required to achieve $<$5\% accuracy across all JWST imaging modes, with a goal to reach $<$1\%--2\% to enable precision science with JWST.  

This paper is part of a broader series detailing the flux calibration of all JWST instruments. In this work, we present the calibration of the Near InfraRed Camera \citep[NIRCam;][]{Rieke+2005, Rieke+2023} imaging modes, including Time Series and Coronagraphy \citep{Girard+2022}. The NIRCam grism calibration is described separately in \citep{Pirzkal+2026}.

\begin{deluxetable*}{llhp{3in}}
\tablecaption{NIRCam Configurations Calibrated in this Work \label{tab:modes}}
\tablehead{\colhead{Mode} & \colhead{Detectors} & &\colhead{Filters/Elements}}
\startdata
Imaging & All 10 detectors& SUB160, SUB160P, SUB64P, FULL & All 29 filters\\
Imaging Time Series & NRCB1 & SUB400P & WLP8 paired with allowed filters \\
Grism Time Series & NRCA1, NRCA3 & SUB320, FULL & WLP4 and WLP8, paired with allowed filters\\
Coronagraphy & NRCA2, NRCA4, NRCALONG & SUB320, SUB640, SUB400X256, FULL & All 5 masks, each paired with allowed filters, plus TA\\
\enddata
\tablecomments{WLP4 and WLP8 are the weak lens elements ($+$4 and $+$8 defocus, respectively). SW detectors are named NRCA1--NRCA4 and NRCB1--NRCB4. LW detectors are named NRCALONG and NRCBLONG.}
\end{deluxetable*}

Because of JWST's high sensitivity, we must extend the flux calibration to fainter targets than those used for previous infrared (IR) missions, while also targeting brighter stars where possible to provide cross-calibration to other missions, such as the Hubble and Spitzer Space Telescopes. The JWST flux calibration program includes observations of three types of stars: A dwarfs, hot stars, and solar analog stars. The previous NIRCam flux calibration (delivered 2023 October) was derived from the Cycle 1 data and included just 3 A dwarfs, 3 hot stars, and 2 solar analogs. Here, we include observations spanning Cycles 1--4 (through 2025 December), which includes 19 stars split between the three types and spanning more than an order of magnitude in flux density. The large number of targets allows for a more robust assessment of the statistical uncertainties and systematic uncertainties between stellar types, and the increased range in flux allows for an assessment of detector effects. Since most standard stars are bright, it is necessary to perform the absolute flux calibration observations using subarrays. We therefore also include an assessment of count-rate offsets between subarrays and the full frame and incorporate it into the flux calibration described in this paper.

\subsection{NIRCam}

NIRCam is the primary near-IR imager on JWST.  There are five NIRCam observing modes: Imaging, Coronagraphy, Time Series (TS) Imaging, Grism TS, and Wide Field Slitless Spectroscopy.  

NIRCam has two modules, A and B, which are essentially two separate instruments with independent optical paths and pupil/filter wheels. This redundancy was built into the instrument because NIRCam is used for wavefront sensing and mirror alignment \citep{Acton+2022}. Data are taken simultaneously in Modules A and B, covering adjacent 2.2\arcmin$\times$2.2\arcmin\ fields on the sky, separated by about 44\arcsec.\footnote{JWST Documentation (JDox) page: \href{https://jwst-docs.stsci.edu/jwst-near-infrared-camera/nircam-instrumentation/nircam-field-of-view\#gsc.tab=0}{NIRCam Field of View}. Note that the reference for all JDox links is \citet{jdox}.}  Each module uses a dichroic positioned before the pupil and filter wheels to split the beam into two channels: a short wavelength (SW) channel that covers 0.6--2.3~$\mu$m, and a long wavelength (LW) channel covering 2.4--5.0~$\mu$m. Data is obtained in both channels simultaneously, covering the same field-of-view in each module.  The SW channel includes 8 detectors (4 per module), and the LW channel includes 2 detectors (1 per module). 

The pupil and filter wheels contain filters, weak lens elements, grisms, and coronagraph Lyot stops.\footnote{JDox page: \href{https://jwst-docs.stsci.edu/jwst-near-infrared-camera/nircam-instrumentation/nircam-pupil-and-filter-wheels}{NIRCam Pupil and Filter Wheels}} Each module includes 13 SW filters and 16 LW filters spanning 4 width categories: narrow (N, $\lambda/\Delta\lambda \approx 78$--92), medium (M, $\lambda/\Delta\lambda \approx 8$--20), wide (W, $\lambda/\Delta\lambda \approx 4$--5), and extra wide (W2, $\lambda/\Delta\lambda \approx 1$--2). The SW channel also includes 2 weak lens elements that defocus the point-spread function (PSF), dispersing it over many pixels and allowing for observations of bright targets. Both channels can be observed with 5 coronagraphic setups on module A, each combining 1 of 5 occulting masks on a substrate in the focal plane with a Lyot stop mounted on a wedge in the pupil plane (1 each for round and bar masks). This paper describes the calibration of all of the elements available for science, except the Dispersed Hartmann Sensing (DHS) elements and the LW grisms (Table~\ref{tab:modes}).  The calibration of the LW grisms is described in \citet{Pirzkal+2026}, and at the time of publication, the DHS elements were still in the process of being commissioned for use with NIRCam's Grism TS mode.


This analysis includes the first on-sky calibration of the weak lenses and the first calibration of the non-primary Coronagraphy channels, which are available as part of the Dual-Channel Coronagraphy capability that was enabled in Cycle 2. We also provide the first subarray-dependent calibrations.  The calibration described in this paper was delivered to the JWST Calibration Reference System (CRDS) and incorporated into the JWST pipeline in 2026 March, as part of pmap 1490.  This calibration converts the images from measured counts (digital number per second; DN s$^{-1}$) to surface brightness units (MJy~sr$^{-1}$), and this conversion is stored in the image headers in the PHOTMJSR keyword after the data are processed through Stage-2 of the JWST pipeline.

In section~\ref{sec:data}, we describe the observations, data reduction, photometry, and subarray count-rate differences. In section~\ref{sec:results}, we discuss the resulting calibration factors for each mode, including residual subarray and detector dependencies, detector-to-detector offsets, and repeatability. In section~\ref{sec:crds}, we describe details of the delivery to CRDS and provide the magnitude zeropoints in the Vega-Sirius system.

\begin{deluxetable}{llcc}[tbp]
\tablecaption{Program IDs\label{tab:pids}}
\tablehead{\colhead{PID} & \colhead{Description} & \colhead{Cycle} & \colhead{Part}}
\startdata
1536 & A dwarf stars & 1 & 1, 2 \\
1537 & Hot stars & 1 & 1, 2 \\
1538 & Solar analog stars & 1 & 1, 2 \\
4496 & A dwarf stars& 2 & 1, 2 \\
4497 & Hot stars & 2 & 1, 2 \\
4498 & Solar analog stars & 2 & 1, 2 \\
6604\tablenotemark{a} & A dwarf stars & 3 & 1, 2 \\
6605 & Hot stars & 3 & 1, 2 \\
6606 & Solar analog stars & 3 & 1, 2 \\
7487 & A dwarf stars & 4 & 1, 2\\
7565\tablenotemark{b} & Hot stars & 4 & 1, 2 \\
7615 & Solar analog stars & 4 & 1, 2\\
\hline
4452 & Full-subarray transfer& 2 & 1 \\
6630 & Full-subarray transfer & 3 & 1 \\
8882\tablenotemark{c} & Full-subarray transfer & 4 & 1 \\
\hline
1539 & Repeatability & 1 & 3 \\
4499 & Repeatability & 2 & 3 \\
6607 & Repeatability & 3 & 3 \\
7671\tablenotemark{c} & Repeatability & 4 & 3 \\
\enddata
\tablenotetext{a}{Program 6604 did not include NIRCam observations, but we include it here for completeness because it is part of the cross-instrument absolute flux program.}
\tablenotetext{b}{Data from program 7565 was not included here because most of it had not been observed at the time of the analysis.}
\tablenotetext{c}{Programs 8882 and 7671 were incomplete at the time of this analysis, but data through 2025~Dec is included.}
\tablecomments{See Tables~\ref{tab:1536}--\ref{tab:8882} for details of each program. } 
\end{deluxetable}

\section{Data}
\label{sec:data}

\subsection{General Program Design}

The overall design of the JWST absolute flux calibration program, including target selection, is described in \citet{Gordon+2022}.  Briefly, the program includes three main parts. Part 1 focuses on the calibration of all observing modes, and observes a calibration star with every filter, optical element, and detector. To mitigate issues with a particular target or target type, we include calibration stars from each of three stellar types: A dwarfs, hot stars, and solar analog stars. Part 2 observes a larger sample of stars with a subset of observing modes to establish the average calibration.  For NIRCam, part 2 includes imaging each target with 5 filters on one longwave (LW) detector and 5 filters on one shortwave (SW) detector. Part 3 measures repeatability by repeatedly observing a calibration star with one filter on every detector. 

This paper includes analysis of data from all three parts of the absolute flux calibration program from JWST Cycles 1--3, and from the first six months of Cycle~4.  Table~\ref{tab:pids} lists the program IDs (PID) included. For parts 1 and 2, the observing programs are organized by stellar type, with each cycle including one program for A~dwarfs, one for hot stars, and one for solar analog stars.  
Table~\ref{tab:targs} summarizes which targets were observed in each mode and Tables~\ref{tab:1536}-\ref{tab:8882} describe the observations for each program. 

\subsection{Observation Description}

Almost all observations were taken using subarrays to avoid saturation.  Exposure times were set to reach a signal-to-noise (S/N) of approximately 200, with some of the narrow filters closer to S/N$\sim$150 for the fainter targets. To reach these S/N requirements, we used readout patterns\footnote{JDox page: \href{https://jwst-docs.stsci.edu/jwst-near-infrared-camera/nircam-instrumentation/nircam-detector-overview/nircam-detector-readout-patterns}{NIRCam Detector Readout Patterns}} that allowed us to achieve the necessary exposure time with $\geq$5 groups, where possible. Multiple integrations were sometimes necessary to reach the required S/N, depending on the target/filter. In some filters, the brightest targets saturated after 2 groups even with the fastest readout patterns. For these, we obtained multiple integrations to decrease the uncertainty in the ramp fit.   

\begin{figure*}
    \centering
    \includegraphics[width=0.85\textwidth]{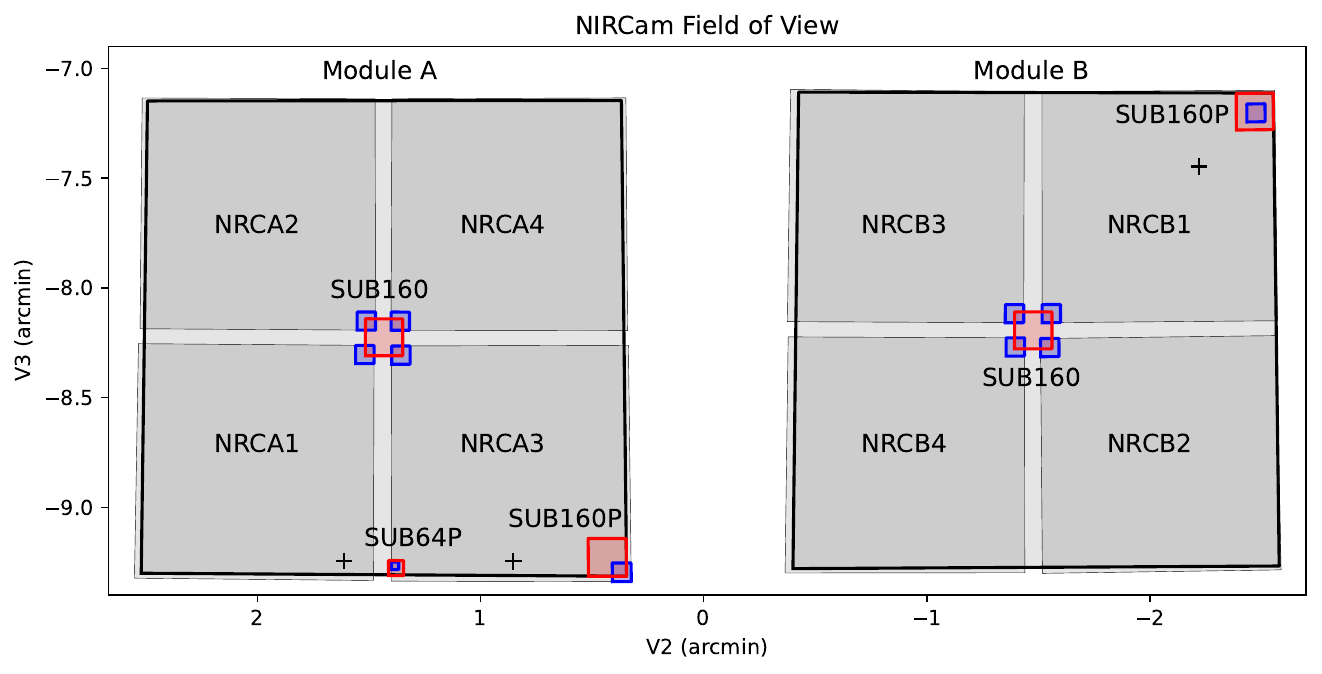}
    \caption{Locations of the primary subarrays used for imaging flux calibration: SUB160, SUB160P, and SUB64P. SW subarrays are blue, LW subarrays are red. SUB64P on Module B is not shown, but is at the same location as SUB160P on that module. The NRCALONG and NRCBLONG detectors are outlined in black. The SW detectors are outlined in gray and labeled.  Note that reference pixels are located in 4-pixel wide boundaries around the edges of each detector, therefore SUB160P/SUB64P on NRCB1, SUB64P on NRCBLONG, and SUB160 on NRCALONG/NRCBLONG do not have reference pixels, resulting in noisier data in those subarrays. The plus symbols mark the locations of the grism time series reference points on Module A and the FULLP time series reference point on module B (The SUB400P time series reference point is the same as SUB160P on module B).  Reference points for the FULL frame are in the center of each detector.  Each of the ten detectors are 2048$\times$2048 pixels.  See the \href{https://jwst-docs.stsci.edu/jwst-near-infrared-camera/nircam-instrumentation/nircam-detector-overview/nircam-detector-subarrays\#gsc.tab=0}{NIRCam Detector Subarrays JDox page} for information on all NIRCam subarrays used for science. Note, the ``P" subarrays on Module A moved to the upper right corner of NRCA4 in 2025 December, after all observations reported here were taken. \label{fig:subs}}
\end{figure*}

\subsubsection{Imaging}

For most imaging observations, we use a combination of the SUB160, SUB160P, and SUB64P subarrays, which are shown in Figure~\ref{fig:subs}.  Note that the subarrays shown on module A are not available for science, but are enabled for calibration purposes.  Also, the SUB64P and SUB160P subarrays on module A were moved to the upper right of NRCA4 in 2025 Dec, after all data analyzed here was collected.

In Cycle 1, we primarily used the SUB160 subarrays (160$\times$160 pixels), which are located at the center of each module. 
To place the star on every SW detector while also keeping it away from the edges of the LW subarray, we used a combination of the INTRAMODULEBOX dither pattern with a 1$\times$2 mosaic. We also included two subpixel dithers so that each SW filter+detector combination was measured at two positions to mitigate bad pixels, cosmic rays, and other artifacts. Since the LW subarray covers all mosaic positions, the star was measured at 8 positions in the LW filters. While this strategy was efficient, the mismatched spatial overlap between the SW and LW channel subarrays made suitable target positioning challenging. The result observations had the target very near the inner corners of the SW detectors (see Fig.~\ref{fig:subs}), where distortions, flat-field effects, and persistence are strongest. 

Starting in Cycle 2, we instead placed the star in the {\em center} of the SUB160 subarray on each SW detector using a 2$\times$2 mosaic. This strategy pushed the star off of the edge of the LW SUB160 subarray (Fig.~\ref{fig:subs}), so we created separate observations for the LW channel. Having a separate LW observation also enabled us to take advantage of reference pixels, which are a 4-pixel wide boundary surrounding the detectors that help with bias and $1/f$ noise corrections. On the LW detectors, SUB160 has no reference pixels because it is at the center of the detector, so we instead use the SUB160P subarrays, which are in the corners of the LW detectors and thus have reference pixels on two sides (Fig.~\ref{fig:subs}). Also starting in Cycle 2, we include four SUBARRAY\_DITHER dithers for both channels.  

The extra-wide filters, F150W2 and F322W2, were sometimes observed on the smaller SUB64P subarrays (64$\times$64 pixels) to avoid saturation of the brightest targets. The SUB64P subarrays are on only two SW detectors (NRCA3 and NRCB1), so the targets that are too bright in F150W2 for the larger SUB160 subarrays are only observed on those two SW detectors.  On Module A, the SUB64P subarrays have reference pixels on one edge, but the Module B the SUB64P subarrays have no reference pixels (Fig.~\ref{fig:subs}), potentially resulting in noisier data.

In Cycle 3, we began including faint white dwarf stars from \citet{Axelrod+2023}, since these stars are faint enough to be observed with NIRCam's full array. We observed these targets in a subset of detectors and filters, placing them near the centers of NRCB1 and NRCA3.

\subsubsection{Time Series}

The Grism TS mode uses the grism in the LW channel, while the SW channel is free for imaging either with the regular imaging filters or with a weak lens.\footnote{Starting in 2026 Jul, there is also the option of obtaining spectra in the SW channel in this mode. See \href{https://jwst-docs.stsci.edu/jwst-near-infrared-camera/nircam-observing-modes/nircam-time-series-observations/nircam-short-wavelength-grism-time-series\#gsc.tab=0}{JDox}.} The Imaging TS mode uses the regular imaging filters and has the option of using weak lenses in the SW channel. The only (non-grism) optical elements specific to the TS modes are therefore the 4$+$ defocus and the 8$+$ defocus weak lenses (WLP4 and WLP8).

The Grism TS mode uses both WLP4 and WLP8, paired with a subset of filters. We include observations of every allowed combination on SW detectors NRCA1 and NRCA3, the only SW detectors used for this mode. We use the FULL frame or the SUBGRISM256 subarrays and place the target at or dithered around the Grism TS reference points (plus symbols in Fig.~\ref{fig:subs}).\footnote{"Reference point" is the terminology used to refer to the default target position within a given subarray/detector.}  The Imaging TS mode uses only WLP8 and only one detector, NRCB1. For that mode, we use the SUB400P subarray (400$\times$400 pixels) for all targets and place the target at the TS subarray reference point (we do not include the FULLP reference point here, which is $\approx$15\arcsec away, see Fig.~\ref{fig:subs}).

\begin{figure}
    \includegraphics[width=\columnwidth]{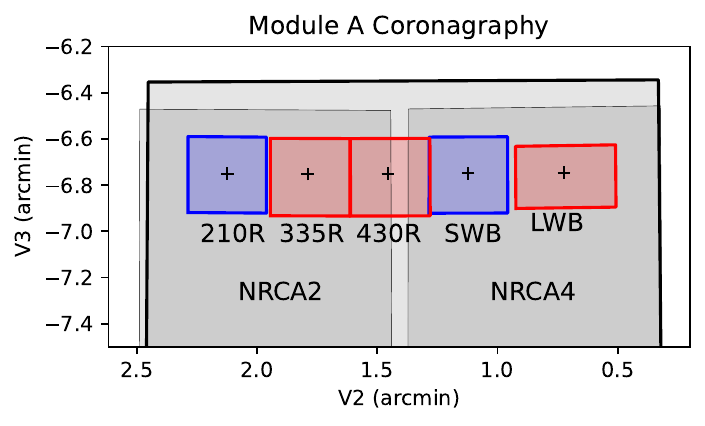}
    \caption{The location of the occulting masks and subarrays for NIRCam Coronagraphy. The plus symbols mark the center of the mask locations. Red and blue outlines indicate the LW and SW subarray on the primary coronagraph channel, respectively. The black line outlines the NRCALONG detector, and the gray lines trace the NRCA2 and NRCA4 SW detectors. SW subarrays are 640$\times$640~pix, LW round mask subarrays are 320$\times$320~pix, and the LWB subarray is 400$\times$256~pix.\label{fig:coron}}
\end{figure}

\subsubsection{Coronagraphy}

NIRCam has 5 coronagraphic masks. The 3 round masks are named for the central wavelength for which they are optimized: 210R, 335R, and 430R. The SW bar mask (SWB) and the LW bar mask (LWB) are optimized for the short and long wavelength channels, respectively. Dual-channel Coronagraphy allows each mask to be paired with both SW and LW filters regardless of the channel they are optimized for. Coronagraphy is enabled only on Module A (Fig.~\ref{fig:coron}).

This analysis includes all the currently available combinations of occulting masks and filters on module A. We note that dual-channel Coronagraphy was not enabled until Cycle 2, so Cycle 1 observations include one channel only for each mask. Depending on the target brightness, we used either the FULL frame or the SUB640 and SUB320 subarrays (640$\times$640 pixels and 320$\times$320 pixels) for the SW and LW channels, respectively. Starting in Cycle 2, SUB320 was replaced by the SUB400X256 subarray (400$\times$256 pixels) for observations using the LWB mask to facilitate dual-channe4499l coronagraphy. The bar masks are observed with the 3-POINT-BAR dither and the round masks with the 5-POINT-BOX dither. In all cases, the target acquisition was performed on the star itself. 

For all coronagraphic observations, we place the star a few arcseconds outside of the mask to ensure that it falls within the coronagraphic optics while remaining unobscured by the mask. This is discussed further in \S\ref{sec:coron_results}.

Starting in Cycle 2, we also began tracking the flux calibration in the target acquisition (TA) subarrays for Coronagraphy, including both the faint-source subarrays (FS) and the bright-source subarrays that include neutral density (ND) filters to suppress the flux. There are currently 7 ND TA subarrays and 5 FS TA subarrays across detectors NRCA2, NRCA4, and NRCA5, located along the bottom of the subarrays shown in Figure~\ref{fig:coron}. TA uses the F210M and F335M filters in the SW and LW channels, respectively. Starting about midway through Cycle 3, NIRCam implemented TA dithering to avoid issues with bad pixels. Prior to that, our TA data consisted of single non-dithered images. The faint-source TA subarrays were used for the general coronagraph absolute flux observations. The ND TA subarrays were included in separate, dedicated observations (see Tables~\ref{tab:1536}--\ref{tab:4452}).

\subsubsection{Repeatability}
\label{sec:repeatdata}

To assess the repeatability, we observe the same target (BD$+$60\ 1753) regularly, through a single SW$+$LW filter pair. These observations are done on SUB160 for both channels in Cycle 1 with two subpixel dithers. Starting in Cycle 2, we moved the LW observations to the SUB160P subarray and increased the number of dithers to $\geq$4.  The observations are described in Table~\ref{tab:1539}.

\begin{figure}
    \includegraphics[width=\columnwidth]{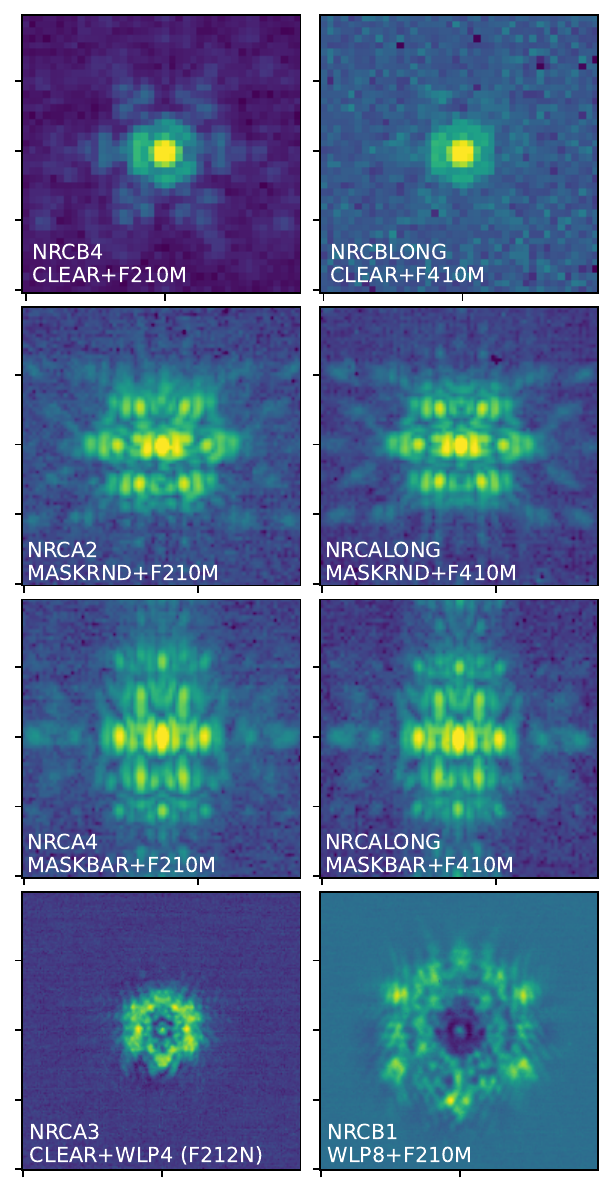}
    \caption{Example PSFs for each type of optical element: imaging filters, coronagraph round and bar masks, and WLP4 and WLP8 weak lenses. All images show F210M or F410M, except WLP4 (F212N).  The top two (imaging) panels are 40$\times$40~pix, the coronagraph panels are 80$\times$80~pix, and the WLP panels are 200$\times$200~pix. The target is P330E in all panels. \label{fig:images}}
\end{figure}

\subsection{Data Reduction and Analysis}


We downloaded the \texttt{*\_uncal.fits} files from the Mikulski Archive for Space Telescopes (MAST) and ran the Stage-1 and Stage-2 pipelines manually using JWST Calibration Pipeline version 1.18.0 and the jwst\_1364 pmap from the Calibration Reference Data System (CRDS), both released in 2025. For Stage-1 (\texttt{CALWEBB\_DETECTOR1}), we increased the threshold for the Jump Detection step in the LW channel, which identifies jumps in the ADU level between groups, usually caused by cosmic rays. Since the LW SUB160 subarrays do not have reference pixels on the edges, the Jump Step occasionally flags too many pixels; increasing the threshold mitigates this issue. We also found several instances (in both channels) where the Jump Detection step mistakenly flagged the core of the target star as an outlier. For these cases, we reran the Stage-1 pipeline with the Jump step turned off. This occurred in about 17\% of the images. 

For both channels, we also turned on the frame0\footnote{See the \href{https://jwst-docs.stsci.edu/jwst-near-infrared-camera/nircam-instrumentation/nircam-detector-overview/nircam-detector-readout-patterns\#gsc.tab=0}{JDox page describing frame0}.} correction in Stage-1 (by setting the suppress\_one\_group argument to False). This allows for saturation that occurs before the end of a group to be corrected 
by performing ramp fitting using the unsaturated frame0 compared to the detector bias level.

To remove $1/f$ noise, we turned on the \texttt{Clean\_Flicker\_Noise} option\footnote{\href{https://jwst-docs.stsci.edu/known-issues/nircam-known-issues/nircam-1-f-noise-removal-methods\#gsc.tab=0}{JDox page} describing the \texttt{Clean\_Flicker\_Noise} step of the Stage-1 pipeline for removal of $1/f$ noise.} in the Stage-1 pipeline, setting background\_method to median and fit\_method to median, and setting the fit\_by\_channel option to False. We then ran Stage-2 (\texttt{CALWEBB\_IMAGE2}) using the default settings. All subsequent analysis uses the resulting \texttt{*\_cal.fits} files. 

In a small subset of images, the exposures are longer than 100 frames. For example, if the readout pattern is DEEP2 with 8 groups, that is the equivalent of 160 frames. Most Dark reference files are currently limited to $\approx$100 frames in the context used here (jwst\_1364.pmap), and therefore no appropriate Dark reference file exists for these images. In these cases, the Dark Subtraction step is automatically skipped in the Stage-1 v1.18.0 pipeline and the noise is therefore slightly underestimated. Because the dark current is low ($<$0.02 DN/s/pix in the LW channel and more than an order of magnitude lower in the SW channel), this has a minimal effect on the data. 

Prior to performing photometry, we converted the \texttt{*\_cal.fits} images into units of DN/s by dividing by the PHOTMJSR keyword found in the image headers. This keyword is the conversion between MJy/sr and DN/s, and it comes from the previous calibration update from 2023 (introduced in jwst\_1126.pmap for imaging and jwst\_1146.pmap for coronagraphy). 
Finally, we multiplied the images by the pixel-area-map to account for the pixel-area dependence that is captured in the flat field due to its being derived from extended sky emission.

Figure~\ref{fig:images} shows example images of the star P330-E from PID 1538 in the F210M (SW) and F410M (LW) filters. 
All targets are well-isolated with no bright stars or other sources visible within the sky apertures used for photometry.  There is no measurable diffuse emission and the backgrounds are therefore flat across the images at all wavelengths. Figure~\ref{fig:images} also shows examples of images taken through the coronagraph optics and with the weak lenses to demonstrate the differences in the point-spread functions (PSFs) compared to imaging.

\subsubsection{Photometry}
\label{sec:phot}

Using the \texttt{*\_cal.fits} images in DN/s, described in the previous section, we performed centroiding and photometry with the python Photutils package, version 2.2.0 \citep{Bradley+2023}. For imaging, the star centroids were located using the DAOStarFinder routine on a cutout image around the expected location of the star. For coronagraphy and ND square TA images, the best results were obtained using the centroid\_quadratic routine.  In both cases, we refined the resulting positions using the centroid\_sources routine with a 2D Gaussian centroiding function.

For weak lens imaging, the PSF frequently does not have a strong central peak, causing centroiding algorithms to fail. Therefore, to find the center of the weak lens PSFs, we first convolved the images with a 2D Tophat filter Kernel with radius 60~pix. Then we used the centroid\_quadratic routine on the resulting convolved image. We visually inspected all images to verify good centroids.

Aperture sizes were set to be large enough to mitigate issues with the brighter-fatter effect (V. Bajaj et al., in preparation) and interpixel capacitance (IPC), while also not being so large as to contain too many bad pixels. Keeping the apertures on the small side is particularly important for the subarrays that do not have reference pixels (SUB160 on NRCALONG/NRCBLONG, SUB64P on NRCB1/NRCBLONG, and SUB160P on NRCB1, see Fig.~\ref{fig:subs}). Subarrays without reference pixels are affected by additional noise, which results in a large fraction of negative pixels that can impact the measured flux if the aperture is too large. The selected aperture sizes are listed in Table~\ref{tab:apersize}. 

The background was subtracted from the aperture flux using a sigma-clipped mean within the background annulus. The uncertainty of the aperture flux was computed by combining the uncertainty returned by the Photutils aperture\_photometry function with the standard deviation within the background annulus.  We experimented with doing a global median background subtraction on each image prior to performing aperture photometry using the Astropy Background2D routine; this resulted in only very marginal changes ($\pm$ $<$0.02\%), so we do not include this step in the final results.

Images with $>$0.5 bad pixels within $r=3$~pix and/or with $>$2\% bad pixels within the entire aperture were excluded from analysis.  The coronagraph and weak lens PSFs are significantly larger and the flux more dispersed than regular imaging (Fig.~\ref{fig:images}), so we relax these restrictions to exclude only images with $>$3 bad pixels within $r=3$~pix and/or $>$10\% bad pixels with the entire aperture.  We also excluded images if any of the aperture fell off the detector edge. We interpolated over bad pixels in the retained images.  

\begin{deluxetable}{llllc}
\tablecaption{Apertures \label{tab:apersize}}
\tablehead{\colhead{Mode/element} & \colhead{$r_{ap}$} & \colhead{$r_{in}$} & \colhead{$r_{out}$} & \colhead{$1/A_{\rm cor}$}\\
\colhead{} & \colhead{(pix)} & \colhead{(pix)} & \colhead{(pix)} & }
\startdata
SW imaging & 5 & 20 & 35 & 0.78--0.82\\
LW imaging & 3 & 20 & 35 & 0.68--0.81\\
SW Coronagraphy & 20 & 70 & 110 & 0.69--0.82\\
LW Coronagraphy & 20 & 50 & 90 & 0.67--0.87\\
SW TA & 20 & 40 & 60 & 0.69--0.78\\
LW TA & 15 & 25 & 60 & 0.68--0.77\\
WLP8 & 80 & 120 & 200 & $\sim$0.96\\
WLP4 & 50 & 80 & 120 & $\sim$0.94
\enddata
\tablecomments{$r_{ap}$ is the aperture radius, $r_{in}$ and $r_{out}$ are the inner and outer radii of the background annulus. The LW TA aperture is smaller than the LW coronagraphy aperture because the LW TA subarray is quite small (64$\times$64~pix). The last column is the resulting fraction of the PSF flux included in the aperture, after correcting for the flux from the PSF wings in the background annulus. Ths inverse of this is the aperture correction, $A_{\rm cor}.$}
\end{deluxetable}

\subsubsection{Aperture Corrections}
\label{sec:apcor}

To extrapolate to an infinite aperture, aperture corrections were derived by computing STPSF simulations for every image \citep[version 2.0.0;][]{webbpsf}. STPSF simulations show very good agreement with the NIRCam PSFs.\footnote{See \href{https://jwst-docs.stsci.edu/jwst-near-infrared-camera/nircam-performance/nircam-point-spread-functions\#gsc.tab=0}{The NIRCam Point Spread Function JDox page}} To configure STPSF, we copied the configuration of each image using the setup\_sim\_to\_match\_file option, which matches the detector, filter, subarray, etc., and downloads the relevant Optical Path Difference (OPD) maps. This ensures that the simulations are an accurate reflection of the state of the mirrors at the time of each observation. We placed the simulated PSF at the same detector position as the star, and for coronagraphy, we used the coron\_shift\_x/y option to move the mask away by the same amount that the star is offset from the mask in the data. The simulated PSFs were oversampled by a factor of 7 and include distortions. For coronagraphy, they are normalized to the exit pupil to remove the effect of the masks. According to these simulations, the apertures used here encompass $\sim$67--87\% of the PSF flux for coronagraphy, $\sim$68--82\% of the PSF flux for imaging, and $\sim$95\% for imaging with the weak lenses, after correcting for the flux from the PSF wings in the background annulus. The aperture correction ($A_{\rm cor}$) is therefore the inverse of this fraction, listed in Table~\ref{tab:apersize}.  

For some dual-channel coronagraphy observations, the star ended up near the edge of the SW subarrays when paired with LW masks. For those observations, we decreased the aperture size, encompassing about 45\% of the PSF flux.

\begin{figure}
    \includegraphics[width=1.1\columnwidth]{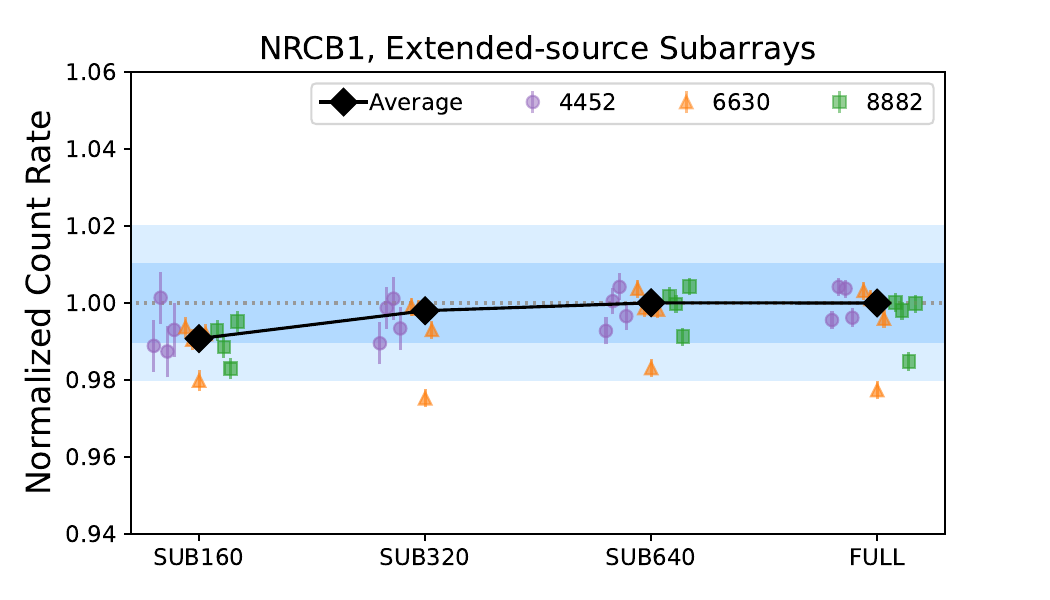}
    \caption{Count-rate differences between subarrays and the full frame. Here, we show the extended source subarrays on detector NRCB1. The complete figure set (29 images) is available in the online journal. In each figure, the star is placed at the same location on the detector for all observations to avoid field dependencies. Data from three programs is included (4452, 6630, and 8882). Each data point represents an individual dither, and is offset in the x-direction to aid in visualization. The black diamonds mark the average. The dark/light blue horizontal bars mark the 1\%/2\% limits. \label{fig:fullsub}}
\end{figure}

\subsubsection{Full Frame vs. Subarrays}
\label{sec:fullsub}

Since most of our flux calibration measurements are done in subarrays, it is necessary to assess whether there are any count-rate differences between measurements made in the full frame to those made in subarrays.  To make this assessment, we observed a standard star on each subarray and at the same detector position in the FULL frame in programs 4452, 6630, and 8882. 
By comparing measurements at the same detector position, we avoid issues with field dependencies. All subarrays used for science or calibration were included. These observations target star C26206 in a single filter pair and 4 dithers using the \texttt{SUBARRAY\_DITHER} pattern.  For Coronagraphy, we use the \texttt{5-POINT-BOX} or \texttt{5-POINT-BAR} patterns. In Cycle 3, we boosted the S/N by increasing the exposure times and switching the coronagraph target to P330E, which is brighter than C26206. 

The grism subarrays each span the entire width of the detector, and therefore span all four amplifiers\footnote{See the \href{https://jwst-docs.stsci.edu/jwst-near-infrared-camera/nircam-instrumentation/nircam-detector-overview/nircam-detector-readout\#gsc.tab=0}{NIRCam Detector Readout JDox page}}, so they have the option of being read out with either one or four outputs. Here, we observed all grism subarrays using 4 outputs ($N_{\rm out} = 4$). All non-grism subarrays are always observed with $N_{\rm out} = 1$. 

Following aperture photometry, we compared the subarray count rates to the corresponding full-frame count rates; the 
results are shown in Figure~\ref{fig:fullsub}. 
To find the subarray offsets, we averaged the count rates measured in each set of dithers with a weighted mean for each program individually, excluding outliers outside 2~sigma. Then, we averaged the results from each program together with a weighted mean.  The Cycle~2 data (PID\ 4452) is shallower than the Cycle~3 and 4 data (PIDs\ 6630 and 8882, respectively), which causes larger uncertainties in Cycle~2 and skews the averages towards the measurements in Cycles 3 and 4.

In general, the smaller subarrays show the largest decreases in the count rate compared to the full frame, up to $\approx$1\%.  We used these offsets to correct all of the absolute flux calibration data that was measured in subarrays to the full frame count-rate before combining measurements taken on different subarrays. 
Subarray-dependent reference files, described in \S\ref{sec:crds}, also ensure that the PHOTMJSR value present in the calibrated image headers includes the appropriate subarray offset. 

We note that we do not have a measured offset for the SUB320 subarray on ALONG used for coronagraphy with the LWB. This subarray was available only in Cycle 1, prior to the implementation of dual-channel Coronagraphy, and was used by just one program: PID\ 1189, for observations of HD\,19467 and its brown dwarf companion, as described in \citet{Greenbaum+2023}. It is likely that this subarray has count-rate offsets similar to the other 320$\times$320 subarrays on ALONG used for MASK210R, MASK335R, MASK430R, and MASKSWB, which are all within $\approx$1\% of the FULL frame.

\begin{deluxetable}{llll}
\tablecaption{CALSPEC2 Models \label{tab:calspec}}
\tablehead{\colhead{Star} & \colhead{Spectral} & \colhead{$K_{\rm s}$} & \colhead{Model} \\
 & \colhead{Type} & \colhead{(mag)} & }
\startdata
\multicolumn{4}{c}{Solar Analog Stars}\\
C26202 & F7V & 14.82 & c26202\_stiswfcnic\_007.fits\\
HR 6538 & G1V & 5.05 & hd159222\_stis\_009.fits\\
P177-D & G0-1V & 11.86 & p177d\_stisnic\_011.fits\\
P330-E & G0V & 11.42 & p330e\_stiswfcnic\_007.fits \\
SNAP-2 & G3V & 14.49 & snap2\_stiswfcnic\_006.fits\\
\hline
\multicolumn{4}{c}{A Dwarf Stars}\\
HR 5467 & A1V & 5.76 & hd128998\_stis\_004.fits\\
J1743045 & A5IIIm & 12.77 & 1743045\_stisnic\_009.fits\\
J1757132 & A8Vm & 11.16 & 1757132\_stiswfc\_006.fits\\
J1802271 & A0V & 11.83 & 1802271\_stiswfcnic\_006.fits\\
J1805292 & A1V  & 12.01 & 1805292\_stisnic\_008.fits\\
\hline
\multicolumn{4}{c}{Hot Stars}\\
10 Lac & O9V & 5.5 & 10lac\_stis\_008.fits\\
G 191-B2B & DA0.8 & 12.76 & g191b2b\_stiswfcnic\_004.fits\\
GD 71 & DA1.5 & 14.12 & gd71\_stiswfcnic\_004.fits\\
GD 153 & DA1.2 & 14.31 & gd153\_stiswfcnic\_004.fits\\
LDS 749B & DB4 & 15.22 & lds749b\_stisnic\_008.fits\\
WD1057 & DA1.2 & 15.47 & wd1057\_719\_stisnic\_011.fits\\
WDFS0122-30 & DA & 20.705\tablenotemark{a} & wdfs0122\_30\_stis\_001.fits\\
WDFS0458-56 & DA & 19.999\tablenotemark{a} & wdfs0458\_56\_stis\_001.fits\\
WDFS2317-29 & DA & 20.423\tablenotemark{a} & wdfs2317\_29\_stis\_001.fits\\
\enddata
\tablecomments{\ Spectral types and $K_{\rm s}$ magnitudes are from \citet{Gordon+2022}, with the exception of the faint white dwarfs (WDFS), which are from \citet{Axelrod+2023}.}
\tablenotetext{a}{We list F160W magnitudes for the faint white dwarfs,  from \citet{Axelrod+2023}.}
\end{deluxetable}

\begin{figure*}
\centering
\includegraphics[width=0.85\textwidth]{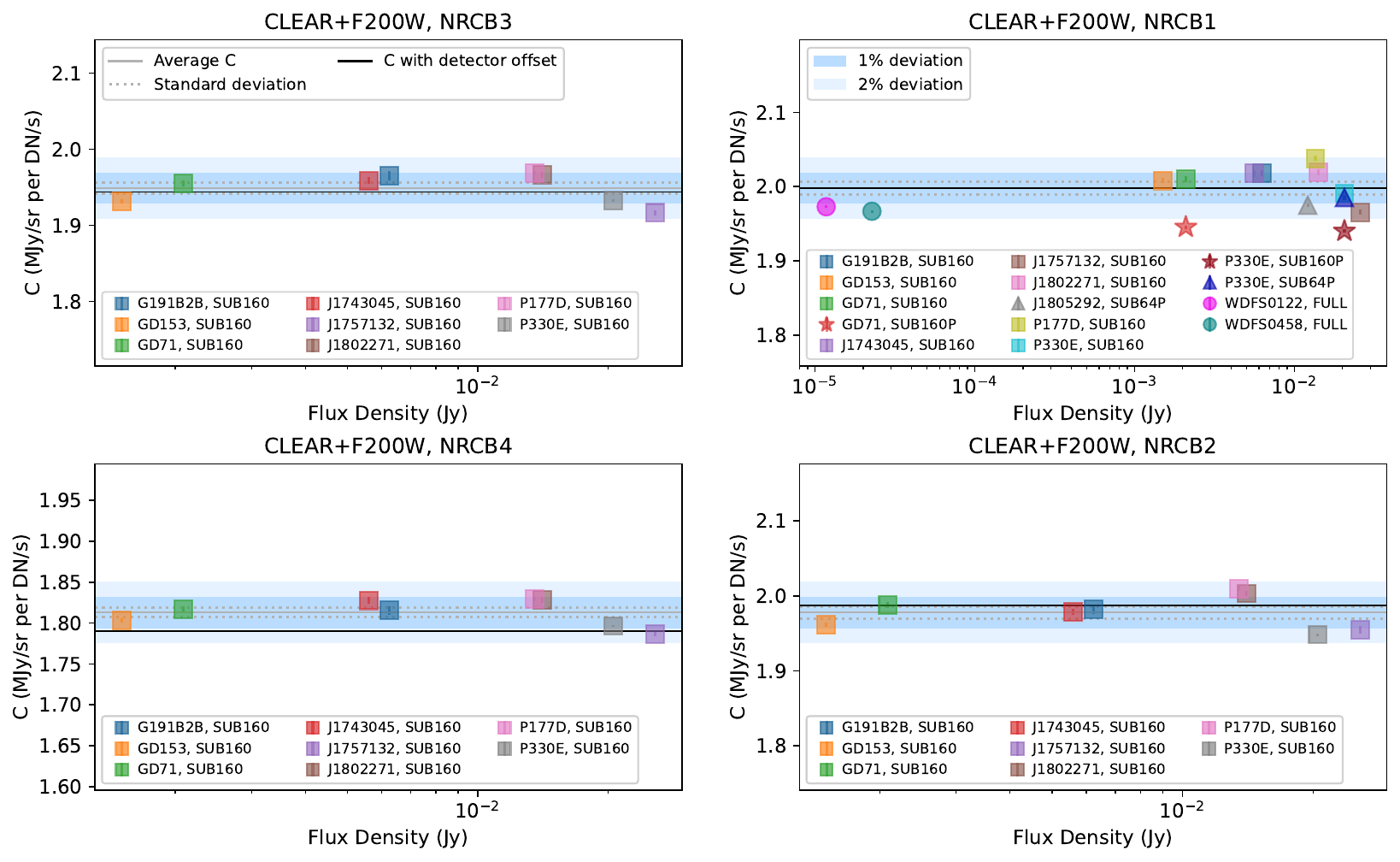}
\caption{The calibration factor ($C$) for F200W on module B, as a function of stellar flux density. The complete figure set (42 images, all imaging filters/detectors) is available in the online journal. Figures are labeled with pupil$+$filter wheel notation (e.g., CLEAR$+$F200W). The panels are arranged as oriented on the sky (see Fig.~\ref{fig:subs}). The gray solid and dotted horizontal lines are the average calibration factor and the standard deviation, respectively. The black line is the calibration factor after detector offsets are applied (\S\ref{sec:detoff}). The dark/light blue horizontal bands mark 1\% and 2\% deviations from the calibration factor. The different symbols
correspond to measurements made in different subar-
rays, which are corrected using the subarray
offsets described in \S\ref{sec:fullsub}.  \label{fig:Cvflux}
}
\end{figure*}

\subsubsection{Model Predictions}


For each standard star, we compare the count rate in DN/s to the predicted stellar flux using models from the CALSPEC2 database \citep{calspec2014, Bohlin+2022}. 
See \citet{Gordon+2022} for the stellar and dust extinction properties of each star. We computed flux densities for every possible combination of optical elements using the Synphot package \citep{synphot} and the version 7.0 filter throughputs, available on JDox.\footnote{Filter throughputs are available on \href{https://jwst-docs.stsci.edu/jwst-near-infrared-camera/nircam-instrumentation/nircam-filters\#gsc.tab=0}{JDox}.}  Table~\ref{tab:calspec} lists the CALSPEC2 models used here.


\subsubsection{Deriving the Calibration Factor}

The flux calibration factor converts instrumental DN~s$^{-1}$~pix$^{-1}$ to surface brightness in units of MJy~sr$^{-1}$. This factor is in the images headers of calibrated images under the keyword PHOTMJSR.  To compute the calibration factors, we follow \citet{Gordon+2022}:

\begin{equation}
C = \frac{F_\nu}{N_{\rm ap}\,A_{\rm cor}\,S_{\rm cor}\,\Omega_{\rm pix}},
\end{equation}

\noindent where $F_\nu$ is the stellar model flux density in MJy, $A_{\rm cor}$ is the aperture correction (\S\ref{sec:apcor}), and $\Omega_{\rm pix}$ is the average solid angle per pixel (sr). $S_{\rm cor}$ is the subarray offset correction described in \S\ref{sec:fullsub}, and $N_{\rm ap}$ is the flux density measured in a finite aperture (DN~s$^{-1}$~pix$^{-1}$), described in \S\ref{sec:phot}.

For each target, we derive average calibration factors (one for each combination of filter and detector) by first combining measurements made on different subarrays with a sigma clipped weighted mean ($\sigma=2.5$) to obtain calibration factors for each target. Then we combine the targets together into a final calibration factor, giving each target equal weight and excluding stars with measurements outside 2.5~$\sigma$.

\section{Results}
\label{sec:results}

In this section, we present the calibration factors for each NIRCam imaging mode: imaging, TS imaging, and coronagraphic imaging. We also describe dependencies, including remaining subarray offsets, differences between stellar types, and trends with the peak count rate, well depth, and with time.

\subsection{Calibration Factors for Imaging}
\label{sec:C_img}

Figure~\ref{fig:Cvflux} shows the calibration factors as a function of the stellar flux density. The different point shapes correspond to measurements made in different subarrays (all points have been corrected using the subarray offsets described in \S\ref{sec:fullsub}). 
The average calibration factor is shown with a gray horizontal line, and two horizontal dotted lines mark the standard deviation. The dark/light blue horizontal bands mark 1\% and 2\% deviations from the mean.  The final calibration factors are listed in Table~\ref{tab:calfacs}. 

Note that detectors NRCA3, NRCB1, NRCALONG, and NRCBLONG have more measurements than other detectors because they include the point-source ``P" subarrays (Fig.~\ref{fig:subs}). The other six detectors include measurements of at least one star per stellar type (A dwarf, hot star, and solar analog) for each filter. The exception is filter F150W2 since many of the standard stars saturate in F150W2 in subarrays larger than our smallest subarray (SUB64P). Therefore, only white dwarf star GD71 was observed in F150W2 on detectors NRCA1, NRCA2, NRCA4, NRCB2, NRCB3, and NRCB4.

\begin{figure}
    \includegraphics[width=\columnwidth]{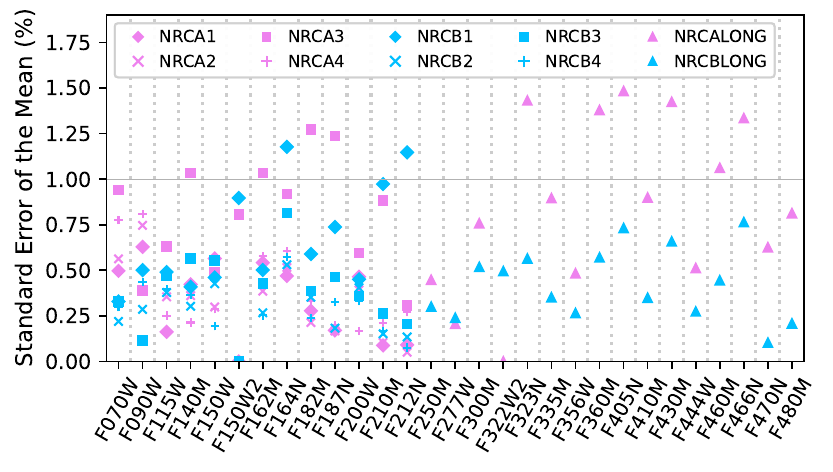}
    \caption{The standard error of the mean (as a percentage) for the calibration factors. The horizontal gray line marks the 1\% level.}
    \label{fig:sem}
\end{figure}

The calibration factors are available in an electronic table that is described in Table~\ref{tab:calfacs}. Overall, most calibration factors for imaging have a standard deviation of $<$2\%, with about half of the filters showing $<$1\% standard deviation. Out of 136 detector$+$filter combinations, 11 show standard deviations between 2\% and 3.2\%, primarily on NRCALONG and NRCA3. For most detector$+$filter combinations, the standard error of the calibration factor mean is $<$1\% (Fig.~\ref{fig:sem}). 

\subsubsection{Anomalous Measurements}
\label{sec:anom}

Most measurements are within 2\% of the mean, with the exception of some measurements of the faint white dwarf stars (WDFS0122-30, WDFS0458-56, and WDFS2317-29) and white dwarf LDS\ 749B. The calibration factor derived for LDS\ 749B is $\approx$2--5\% higher than the average at $\lambda \gtrsim 2~\mu$m.  For WDFS2317, the calibration factor is up to $\sim$5\% too low in the LW channel.  WDFS0458-56 and WDFS0122-30 are both mostly in agreement with other measurements, but are discrepant by up to $\sim$2\% in F070W and some LW filters.  

The JWST/NIRISS absolute flux analysis \citep{Volk+2025} did not include the faint white dwarfs, but did find discrepant measurements for LDS\ 749B, up to 9\% for the F277W to F480M filters. The JWST/MIRI absolute flux observations did not include any of these stars \citep{Gordon+2025}, but that analysis did find some discrepancies for solar analog C26202 at some MIRI wavelengths. C26202 does not appear to be a notable outlier in any NIRCam detectors/filters.

The discrepant measurements for LDS\ 749B and WDFS2317 suggest there may be issues with the CALSPEC2 models. WDFS2317 is type DA, which have fully radiative pure hydrogen atmospheres that should be straightforward to model. LDS\ 749B is a type DB white dwarf, and therefore has a helium dominated atmosphere.

We exclude all three faint white dwarfs and LDS\ 749B from the average calibration factor for all filters/detectors, but include them on the plots for comparison (Fig.~\ref{fig:Cvflux}, and 
Figs.~\ref{fig:peak} and \ref{fig:well} in \S\ref{sec:det}). Additional faint white dwarfs selected from \citet{Axelrod+2023}, WDFS1302+10 and WDFS1557+55, are also included in our observing programs, but had not yet been observed at the time of this analysis. Faint stars like these white dwarfs are desirable targets for absolute flux analyses because they do not saturate in NIRCam full-frame imaging.

\begin{deluxetable*}{crrrrrrr|cr}
\tablecaption{Detector-to-Detector Magnitude Offsets \label{tab:offsets}}
\tablehead{\colhead{SW Filter} & \colhead{NRCA1} & \colhead{NRCA2} & \colhead{NRCA3} & \colhead{NRCA4} & \colhead{NRCB2} & \colhead{NRCB3} & \multicolumn{1}{c|}{NRCB4} & \colhead{LW Filter} & \colhead{NRCALONG}}
\startdata
F070W & $-$0.017 & $-$0.006 & $-$0.055 & $-$0.032 & $-$0.007 & 0.004 & $-$0.028 & F277W & $-$0.002\\ 
F150W & $-$0.002 & $-$0.006 & $-$0.028 & 0.003 & $-$0.005 & $-$0.002 & 0.014 & F356W & $-$0.024\\ 
F200W & 0.006 & 0.007 & $-$0.038 & 0.002 & $-$0.005 & 0.003 & 0.014 & F444W & 0.001\\ 
F210M & 0.023 & 0.022 & $-$0.022 & 0.023 & 0.009 & 0.011 & 0.024 & F335M & $-$0.007\\ 
\enddata
\tablecomments{These offsets are measured in magnitudes, and the uncertainty on each is 0.001~mag. Offsets are measured relative to detector NRCB1 in the SW channel and NRCBLONG in the LW channel.}
\end{deluxetable*}

\begin{figure}
\centering
    \includegraphics[width=0.9\columnwidth]{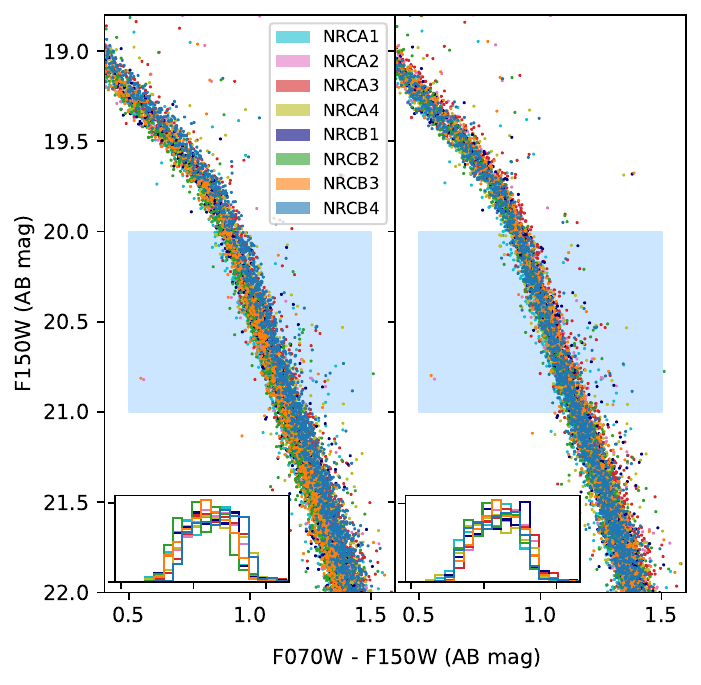}
    \caption{Color-magnitude diagram zoomed into the main sequence of 47\,Tuc. Left and right panels show before and after detector offsets are applied, respectively.  Inset histograms show the color distribution of stars measured on each detector in the blue shaded region and more clearly show the differences between detectors. Histograms are binned by color with a bin size of 0.03~mag.  
    \label{fig:cluster}
    }
\end{figure}

\subsubsection{Residual Detector Offsets}
\label{sec:detoff}

We used observations of the Large Magellanic Cloud (LMC) and globular cluster 47\,Tuc to assess whether any flux offsets between detectors persist after applying the new calibration factors. Cluster data is particularly useful for visualizing detector offsets because shifts in the calibration between detectors cause a broadened main sequence \citep[or even multiple main sequences;][]{Boyer+2022}.  The 47\,Tuc data are from calibration program 6631, and the LMC observations are from commissioning and calibration programs 1069 and 1476. We ran the data through the standard JWST pipeline, using the new calibration factors derived in this work. 

We then produced PSF photometry using the \texttt{1pass} code \citep{Anderson+2022, Bajaj_One-Pass}, which was originally designed for Hubble data and has been adapted for use with JWST.  To remove bad pixels, diffraction spikes, contaminated sources, etc., from the catalog, we cut sources with a PSF fit-quality metric ($q$) that falls within the worst 10\% of all detected sources. The $q$ metric is defined as the residual PSF fit divided by the stellar flux: 

\begin{equation}
q = \sum_{\rm aperture} |P_{\rm ij} - z_*\Psi_{\rm ij} - s_*|/z_*
\end{equation}

\noindent where $P_{\rm ij}$ is the pixel value, $z_*$ is the stellar flux, $\Psi_{\rm ij}$ is the predicted fraction of the star's flux in that pixel, and $s_*$ is the measured sky value. We show the 47\,Tuc data in the left panel of Figure~\ref{fig:cluster} for filters F070W and F150W. A slight broadening of the main sequence caused by detector flux offsets is apparent.  

The LMC observations can be used to derive these detector offsets since they were mosaicked in a way that made all 10 detectors completely overlap one another on they sky. Since the same set of stars was observed on each detector, we can directly compare the magnitudes between detectors and compute the mean magnitude offsets. These observations were taken in a handful of filters: F070W, F150W, F200W, and F210M in the SW channel and F277W, F335M, F356W, and F444W in the LW channel. To measure the detector offsets for these eight filters, we restricted the catalog to 19--21~mag in each filter in the SW channel and 20--22~mag in the LW channel. These magnitude ranges are about 2 magnitudes fainter than the saturation limit, while also remaining bright enough to avoid measurements with large photometric uncertainties. Each offset was calculated using a sigma clipped mean ($\sigma = 2.5$), leaving $\approx$400--3200 stars distributed evenly across each detector with average magnitude uncertainties of 0.003--0.01~mag.

We find the average offsets (measured relative to detectors NRCB1 and NRCBLONG) are typically $<$0.01~mag, or $\approx$1\%, with the largest offset of 0.055~mag for NRCA3 in F070W. The right panel of Figure~\ref{fig:cluster} shows the improved 47\,Tuc data with these offsets applied.  We list all measured offsets in Table~\ref{tab:offsets}, and have applied them to the calibration factors for these eight filters in the reference files delivered to CRDS (see \S\ref{sec:crds}). Figure~\ref{fig:Cvflux} includes black horizontal lines marking the offset calibration factors for the filters with offset measurements.

While we have only four filters in each channel to assess here, it is worth noting that detectors NRCA3 and NRCB4 show the largest offsets (measured relative to NRCB1), ranging from 0.014--0.028~mag for NRCB4 and 0.022--0.055~mag for NRCA3. It is reasonable to assume that these detectors may have the largest offsets in every SW filter.  In the four LW filters measured here, the offsets between NRCALONG and NRCBLONG are generally quite small ($<$0.01~mag), with F356W showing the largest offset at 0.024~mag. It is unclear from this information how large the offsets might be in other LW filters. We stress that the calibration factors delivered to CRDS include measured offsets only for the eight filters listed in Table~\ref{tab:offsets}. A future Cycle~5 calibration program will measure detector offsets in every filter (PID 12538).

\begin{figure}
\centering
    \includegraphics[width=0.85\columnwidth]{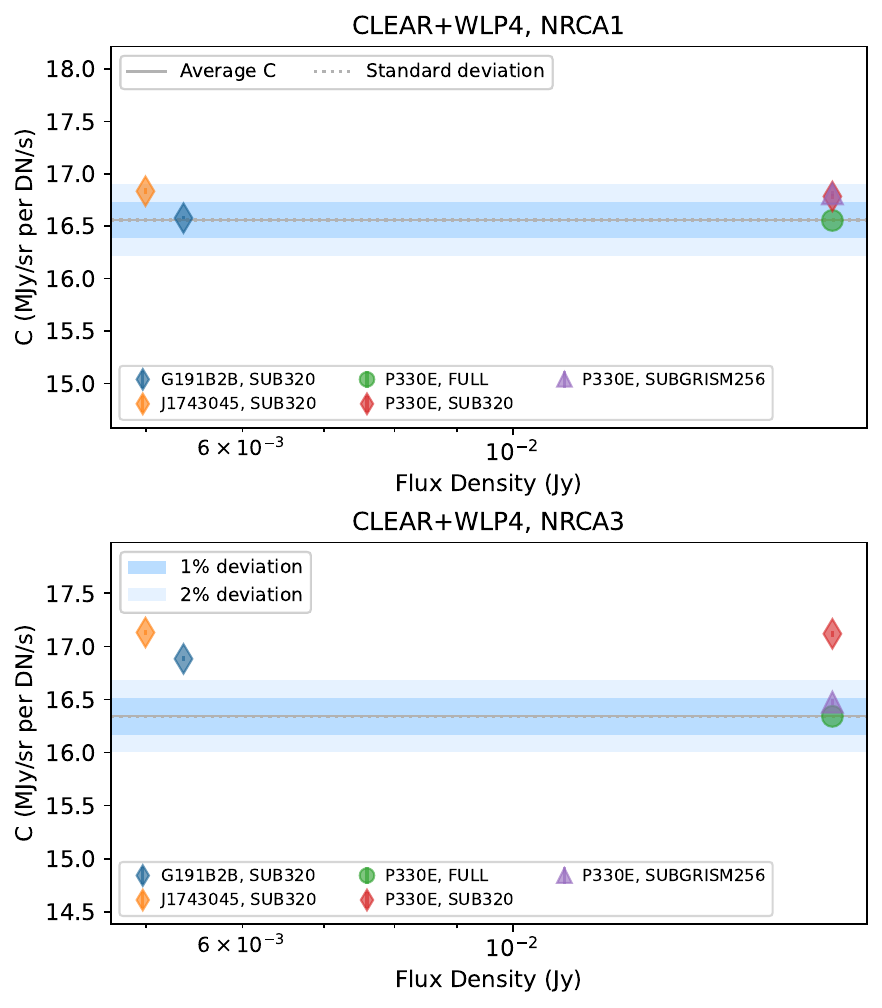}
    \caption{Same as Figure~\ref{fig:Cvflux}, but for the weak lens elements: WLP4 and WLP8. Here we show CLEAR+WLP4. The complete figure set (14 images, all WLPs/filters/detectors) is available in the online journal. We do not include plots for WLP8+F200W on NRCA1 and NRCA3 because only one star was measured in those cases.\label{fig:C_wlp}}
\end{figure}

\subsection{Calibration Factors for Time Series Imaging} 

The NIRCam Grism Time Series mode (module A) uses weak lens elements (WLP4 and WLP8) for imaging in the SW channel. The Imaging Time Series mode (module B) allows the use of WLP8 or regular imaging filters in the SW channel and just regular filters in the LW channel.  This analysis provides the first in-flight flux calibration for WLP4 and WLP8. Figure~\ref{fig:C_wlp} shows how the calibration factor varies with source flux for all weak lens filter+detector combinations. 

\subsubsection{Weak Lens Elements}

The weak lenses can be used only for time series observations, either on detectors NRCA1 and NRCA3 in the Grism Time Series observing mode, or on detector NRCB1 in the Imaging Time Series observing mode.  Because these are time series modes, the target is intended to always be at the same detector location with no dithering. Therefore, the weak lens observations are also at these same detector locations to avoid introducing any field dependencies (see plus symbols in Fig.~\ref{fig:subs}). However, in Cycle~1, we included additional data in the SUB320 subarrays on Module A, well away from the science reference points.  We include these data in Figure~\ref{fig:C_wlp}, where it is apparent that flat field effects are as high as 3-4\% with the weak lenses (especially on NRCA3). This is presumably because the weak lens flat fields are still based on ground test data at the time of this analysis, which are known to vary by up to 6--8\% compared to sky flats \citep{Sunnquist+2022}. We exclude the SUB320 data from our final weak lens calibration factor averages and include only data taken at (and dithered around) the relevant science reference point.

On Module B, the weak lens absolute flux data were taken using the SUB400P subarrays. 
Note that the reference point for SUB400P is the same for subarrays SUB160P and SUB64P, but it is about 15\arcsec\ away from the FULLP reference point used for full-frame Time Series Imaging (see Fig.~\ref{fig:subs}). The programs analyzed here did not include weak lens measurements at the FULLP location, and it is possible that FULLP science measurements will be slightly offset from those taken in the subarrays due to weak lens flat field uncertainties. Cycle~5 calibration observations will include this FULLP reference point. 

Because we are not including observations taken on SUB320, only the solar analog star P330E is included in the final WLP4/WLP8 calibration factors on Module A. The final WLP8 calibration factors for Module B also include A dwarf star J1743045 and white dwarf G\,191-B2B. More targets will be included in Cycle~5. The final calibration factors are listed in Table~{\ref{tab:calfacs}}. 

\subsubsection{TS Standard Imaging Filters}

The calibration factors described in \S\ref{sec:C_img} for standard imaging filters are also applied to the Imaging TS mode on module B (the Grism TS mode only uses weak lenses and grisms). These values include data from the point-source ``P" subarrays, which are at the TS field points. However, the SUB160 subarray on the opposite corner of the NRCB1 detector (Fig.~\ref{fig:subs}) has more measurements and thus dominates the averages (see Fig.~\ref{fig:Cvflux}). In section \S\ref{sec:sub}, we discuss residual subarray offsets, which may affect imaging TS measurements.

\begin{figure}
\centering
\begin{subfigure}{\columnwidth}
    \centering
    \includegraphics[width=0.85\textwidth]{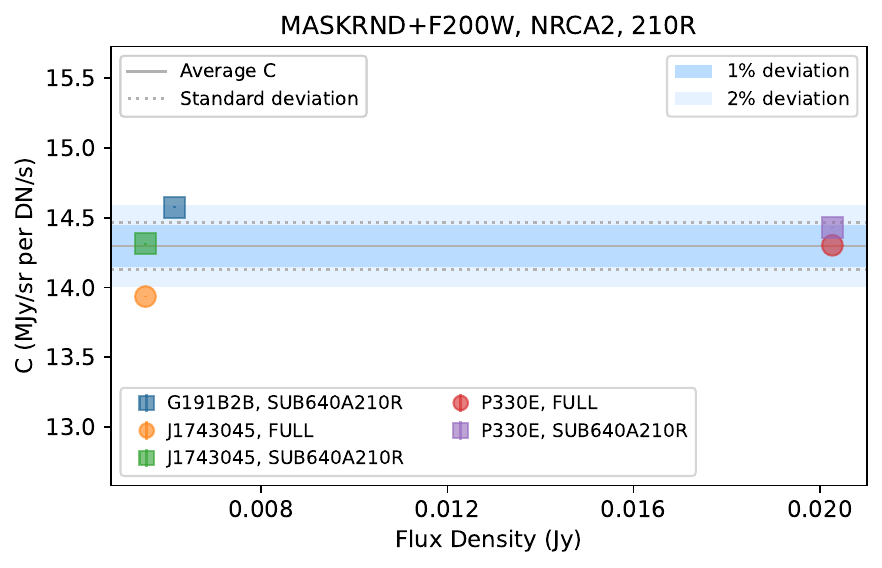}
\end{subfigure}
\begin{subfigure}{\columnwidth}
    \centering
    \includegraphics[width=0.85\textwidth]{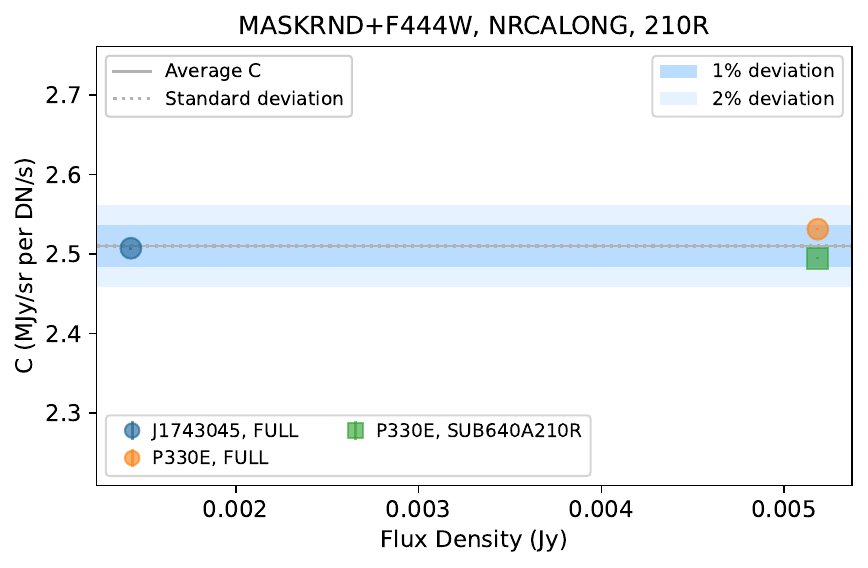}
\end{subfigure}
\caption{Same as Figure~\ref{fig:Cvflux}, but for coronagraphic imaging. Here we show F200W and F444W for the 210R mask. The complete figure set (48 images, all filters/masks/detectors) is available in the online journal. We do not include plots with only one measurement. 
\label{fig:C_coron}}
\end{figure}

\begin{figure}
    \centering
    \includegraphics[width=0.9\columnwidth]{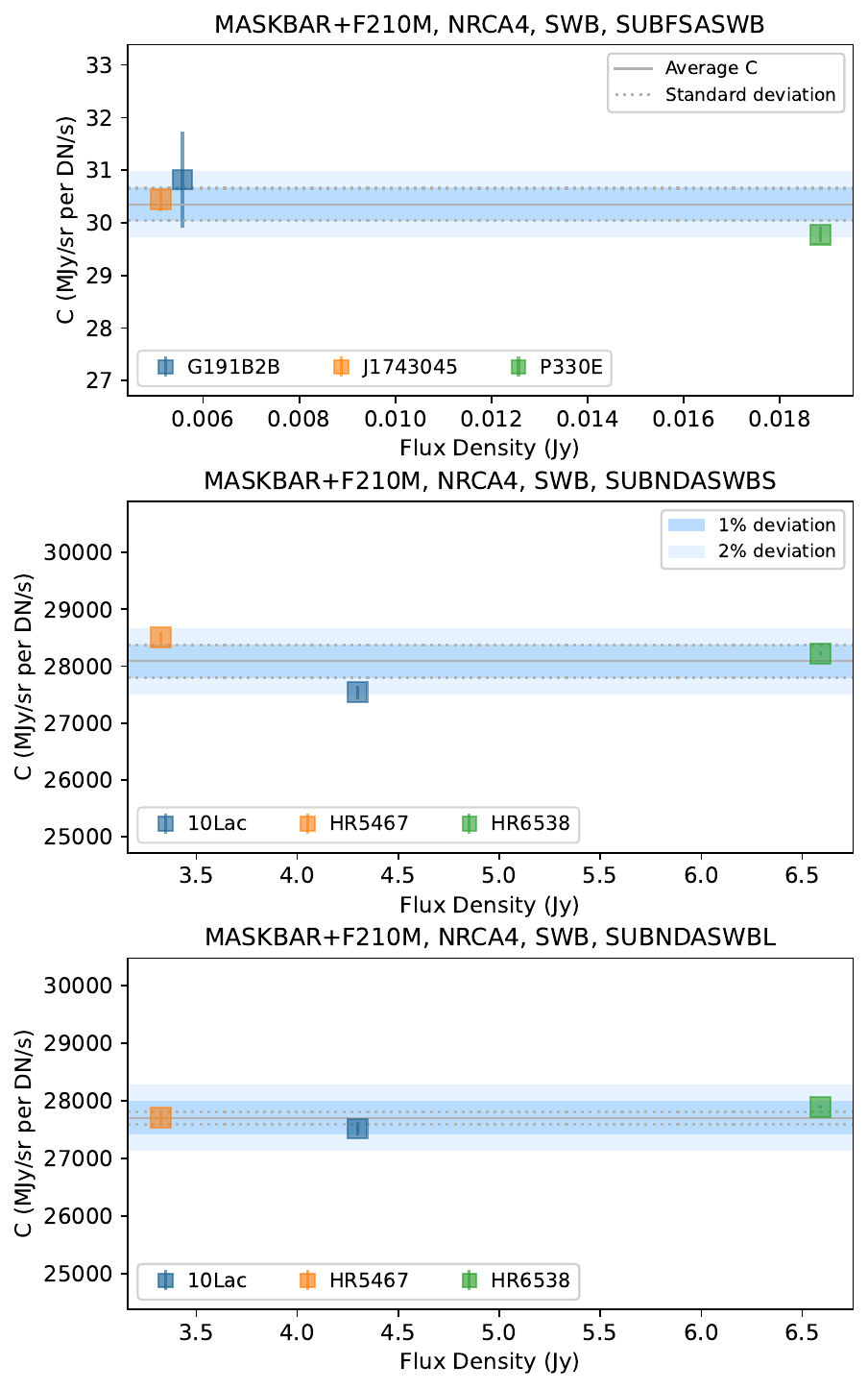}
    \caption{Same as Figure~\ref{fig:Cvflux}, but for coronagraphic target acquisition. Here we show the three TA subarrays for the SWB: one faint-source subarray (SUBFSASWB) and two neutral density subarrays (SUBNDASWBS and SUBNDASWBL). The complete figure set (5 images) is available in the online journal. \label{fig:C_ta}}
\end{figure}

\subsection{Calibration Factors for Coronagraphy}
\label{sec:coron_results}

The coronagraph calibrations factors are plotted as a function of stellar flux in Figure~\ref{fig:C_coron} and listed in Table~\ref{tab:calfacs}. 

In Cycle~1, the position angles of the coronagraphic observations were not properly constrained, causing the LWB mask to partially obscure the star in all three programs. This was made worse by the uncertain positions of the LWB center in Cycle~1. The stars in Cycle~1 are therefore only about 0\farcs3 -- 1\farcs1 from the bar center in the detector y-direction, but the LWB transmission affects objects within about 2\arcsec\ of the bar center at its widest point \citep[see Fig.~1 from ][]{Balmer+2025}. The flux calibration delivered in 2023 Oct (pmap 1146) includes the two stars that are the farthest from the LWB center, but these results appear to disagree with the Cycle 2--4 data by 5\%--10\% in most filters. We therefore exclude all LWB data from the Cycle-1 programs from this analysis. 

Similarly, stars must be $\gtrsim$0.5\arcsec\ away from the center of the SWB mask in the detector y-direction to avoid being affected by its transmission. Star G\,191-B2B in program 1537 does not meet this criterium, so we exclude it from the analysis.  We retain the other two stars from Cycle~1 (P330E in PID~1538 and J1743045 in PID~1537) since they are $>$1\arcsec\ away from the SWB center. 

Most coronagraph observations were taken in the subarray associated with each mask (see Fig.~\ref{fig:coron}), though some Cycle~4 data were taken in the full frame, allowing the star to be farther from the mask.  In program 7487, star J1743045 was placed $\approx$15.5\arcsec\ away from the SWB mask, or $\sim$500~pix in the SW channel and 250~pix in the LW channel.  We find that this is a large enough distance for flat field effects to come into play in the channel.  At the time of this analysis, the coronagraph flat fields still rely on ground-based data and are not well constrained. As a result, the data for this observation is $>$5-10\% offset from the other observations and we therefore exclude it from the average. 
The remaining observations are all near enough to the mask to avoid field dependencies larger than 1--2\%, yet far enough to avoid being affected by the mask's transmission.

Because we exclude some Cycle 1 data and because dual-channel coronagraphy was not enabled until Cycle 2, there are many coronagraph mask$+$filter combinations that have not yet been observed with all three stellar types (hot stars, A dwarf, and solar analogs).  The primary channels for the round masks (mask 210R in the SW channel and masks 335R and 430R in the LW channel) are the only combinations where the calibration factor averages include all three stellar types. The secondary channel combinations (SW masks paired with LW filters and vice versa) include only solar analog P330E, with the exception of mask 210R, which also includes A dwarf star J1743045 in the LW filters. The LWB mask has no measurements with SW filters because the target was too close to the subarray edge for accurate photometry. We discuss how we handle the CRDS delivery for the LWB in \S\ref{sec:crds_notes}.

\subsubsection{Target Acquisition}

This analysis includes the first in-flight calibration of the coronagraph TA subarrays, providing users with an opportunity to obtain an absolute flux measurement of their unocculted target.  Figure~\ref{fig:C_ta} shows the calibration factors for the TA subarrays as a function of stellar flux density. The faint-source subarray data (prefixed with ``SUBFS") have large error bars due to low signal-to-noise. The data on ND TA subarrays (prefixed with ``SUBND") have higher signal-to-noise. In both cases, there are no clear trends with source flux density. Table~\ref{tab:calfacs} lists the calibration factors.

\begin{figure}
\centering
\begin{subfigure}{\columnwidth}
    \centering
    \includegraphics[width=0.85\textwidth]{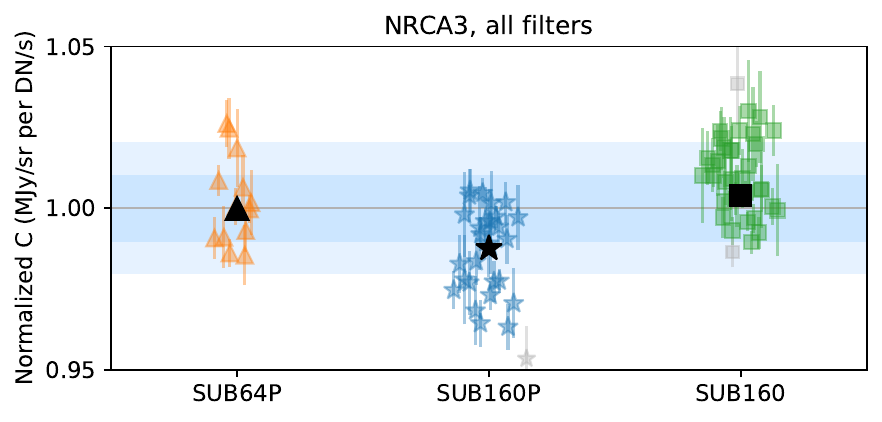}
\end{subfigure}
\begin{subfigure}{\columnwidth}
    \centering
    \includegraphics[width=0.85\textwidth]{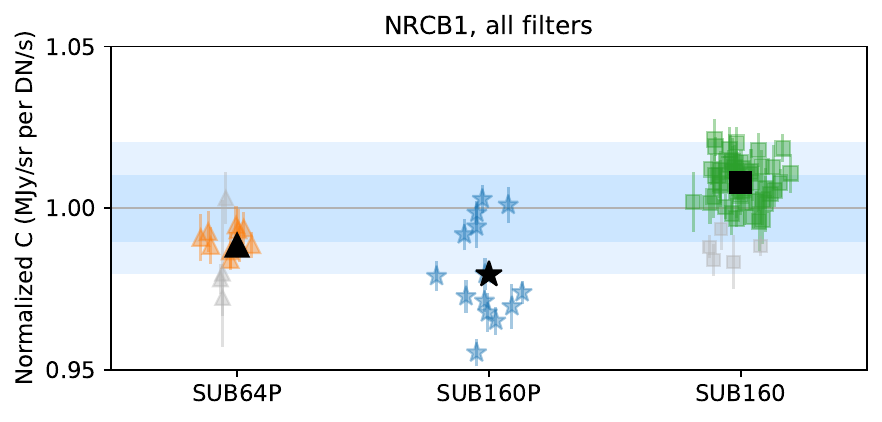}
\end{subfigure}
\begin{subfigure}{\columnwidth}
    \centering
    \includegraphics[width=0.85\textwidth]{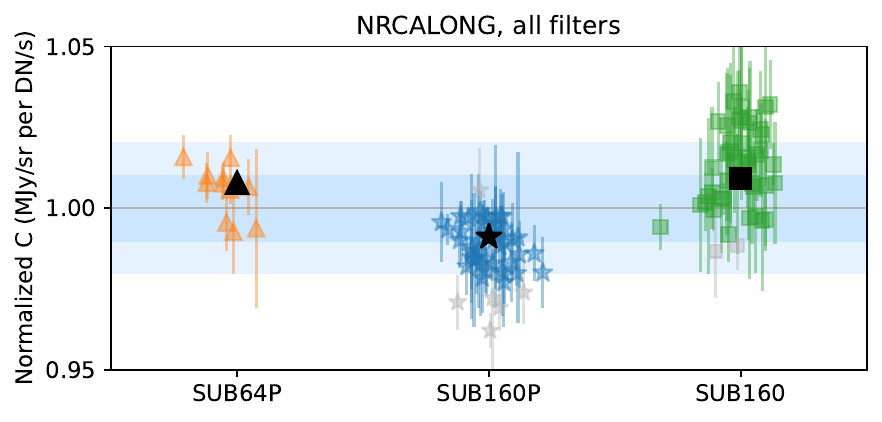}
\end{subfigure}
\begin{subfigure}{\columnwidth}
    \centering
    \includegraphics[width=0.85\textwidth]{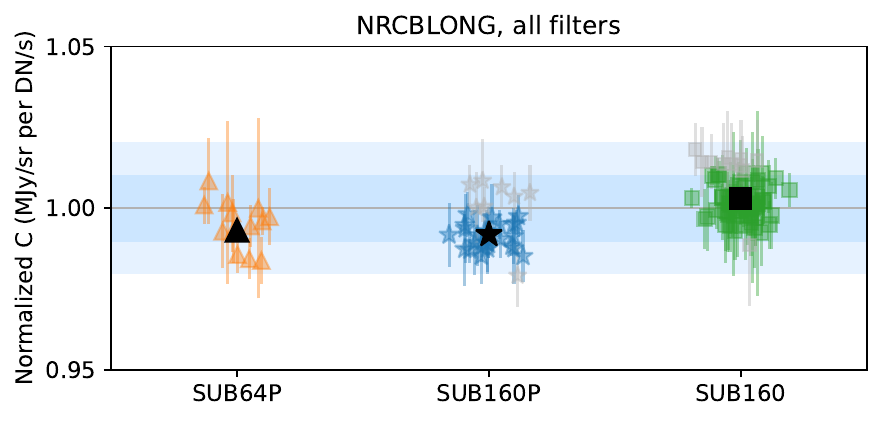}
\end{subfigure}
\caption{The calibration factor vs. the subarray, on the four detectors where measurements were made on multiple subarrays. We include all filters in each panel, with the calibration factor normalized to the average in that filter. Random scatter to the left and right has been added to aid in visualization. The horizontal dark/light blue bars mark the 1\%/2\% boundaries. The black points are the average for each subarray.  3-$\sigma$ error bars are plotted. Gray points are outside 3-$\sigma$ and are not included in the averages.
\label{fig:calfac_sub}}
\end{figure}

\subsection{Calibration Factor Dependencies}

\subsubsection{Subarray Dependencies}
\label{sec:sub}

To assess the subarray offsets computed in \S\ref{sec:fullsub}, we show the calibration factors, corrected to the FULL frame, as a function of the subarray in Figure~\ref{fig:calfac_sub} for the detectors with measurements in multiple subarrays: SUB64P, SUB160P, and SUB160.  All standard imaging filters are included in each panel, and the averages for each subarray (black points) exclude points outside 3-$\sigma$ (gray points).   

The resulting averages are within about $\pm$1\% for the NRCA3 and the LW detectors. 
These small residuals might be explained by lingering flat field effects, since subarrays are located in many different parts of the detectors, often in the corners where the flat field can be difficult to characterize (Fig.~\ref{fig:subs}). This premise is supported by the residuals on NRCBLONG, where SUB64P and SUB160P (which are located at the same spot on the detector) show similar residuals, but SUB160 is offset. On NRCA3 and NRCALONG, the three subarrays are dispersed across the detector, and each shows a different offset. 

The subarrays on NRCB1 show quite large residuals, especially between SUB160P and SUB160, which differ by $\sim$3\%.  However, the two ``P" subarrays, which are at the same location on NRCB1 (Fig.~\ref{fig:subs}) agree within $\sim$1\%, again suggesting that flat field effects may be the cause of the SUB160 offset. The average is biased towards SUB160, which has the most measurements, by far. 

Currently, the NIRCam flat fields for imaging are derived from diffuse sky emission, wherein many images are stacked to remove stars \citep{Sunnquist+2024}.  The NIRCam team is in the process of deriving and analyzing new `stellar' flat fields, derived by stepping stars across each detector. Preliminary results show that the new stellar flats decrease the offsets between the subarrays and the full frame (\S\ref{sec:fullsub}), and decrease the scatter in the calibration factors. Analysis is still underway (B. Sunnquist et al., in preparation).

Another possibility is that the subarray offsets derived in \S\ref{sec:fullsub} are not accurate. For SUB160 and SUB160P on NRCALONG, there is significant scatter (Fig.~\ref{fig:fullsub}), which may be the cause of the residual discrepancy in Fig.~\ref{fig:calfac_sub} for that subarray. The subarray offsets for NRCB1 also show scatter that is large enough to explain the residual discrepancies on NRCB1. Future calibration observations will obtain deeper exposures and more dithers to reduce the scatter and produce more precise subarray offsets.

Many filter+detector combinations have measurements in multiple subarrays, enabling a robust average. However, the SW point-source ``P" subarrays are located {\em only} on detectors NRCB1 and NRCA3, and the other six SW detectors are restricted to measurements on SUB160 (Fig~\ref{fig:subs}). A bias in the SUB160 subarrays on these six detectors likely contributes to the SW detector offsets discussed in \S\ref{sec:detoff}.

We note that the residual subarray offsets on NRCB1 and NRCBLONG will affect Time Series Imaging observations, which use the point-source subarrays on those detectors. Since the calibration factors are biased towards the SUB160 subarrays, Time Series observations that use the standard imaging filters may be discrepant by up to about 2\% on NRCB1.

Plots similar to Figure~\ref{fig:calfac_sub} for coronagraph and WLP8 subarrays are not useful since the final calibration factors for those modes do not average measurements on multiple subarrays. For coronagraphy, we deliver separate calibration factors for each mask/subarray. For WLP8, only one subarray is used here (SUB400P).  For WLP4, only SUBGRISM256 is used. 

\begin{figure}[h!]
\subfloat{%
\includegraphics[width=0.97\columnwidth]{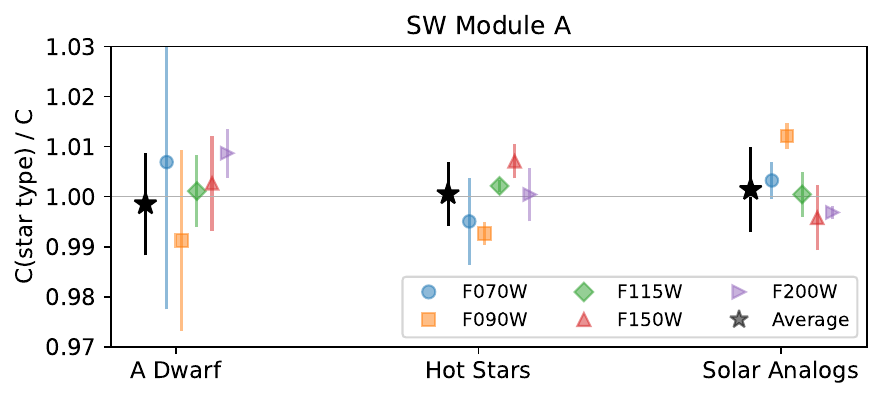}
}\\ 
\subfloat{%
\includegraphics[width=0.97\columnwidth]{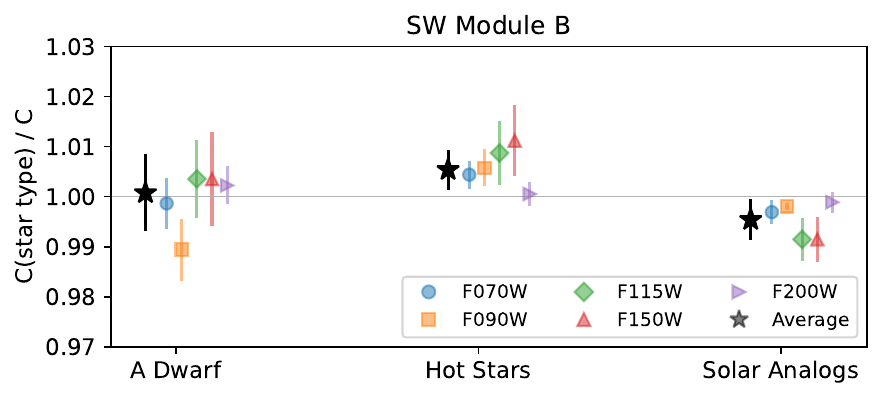}
}\\ 
\subfloat{%
\includegraphics[width=\columnwidth]{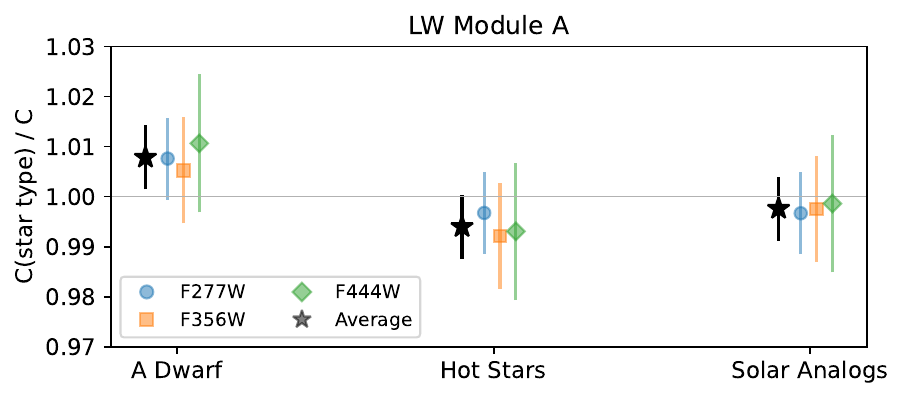}
}\\
\subfloat{%
\includegraphics[width=\columnwidth]{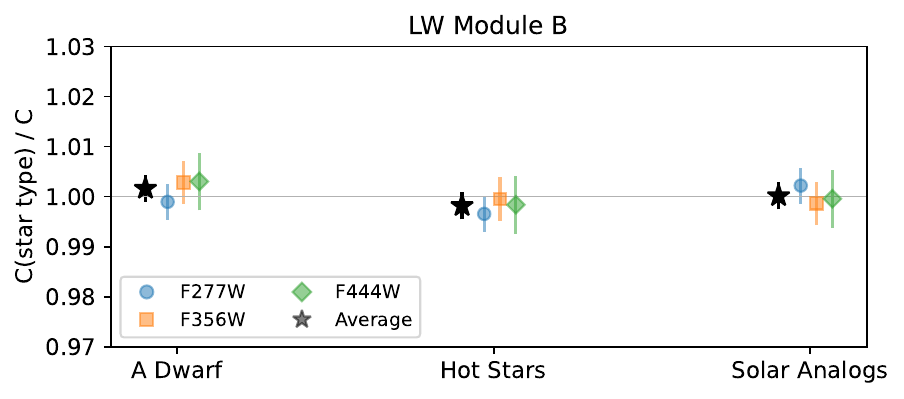}
}
\caption{The dependence of the calibration factor on the source type. The SW detectors are combined into a single panel for each module. 3-$\sigma$ error bars are included. The black points are the average of all wide filters, and the colored points are the individual filters. No trends are visible for any channel/module. \label{fig:avgcal}}
\end{figure}

\subsubsection{Source Dependencies}

In Figure~\ref{fig:avgcal}, we show how the average calibration computed for each of the wide filters varies with source type. This was computed by normalizing the average calibration factor for a given filter for each source type to the average calibration factor in that filter for all three source types, excluding the stars discussed in \S\ref{sec:anom}. Overall, all three source types (A~dwarf, hot star, and solar analog) agree with the average within 3-$\sigma$ and the scatter is $\lesssim$1\%.

A~dwarf stars appear elevated compared to the other two stellar types in the LW channel of Module~A.  This is likely connected to the residual subarray offsets seen in the SUB160 subarray on that detector (\S\ref{sec:sub}). Star J1743045 was observed in that subarray and is elevated compared to the average in all three wide filters by $\approx$2\%, albeit with large uncertainties. Since only one other A~dwarf star is included in the average for that detector in Figure~\ref{fig:subs} (J175132), the average is slightly elevated.

\begin{figure}
    \centering
    \includegraphics[width=\columnwidth]{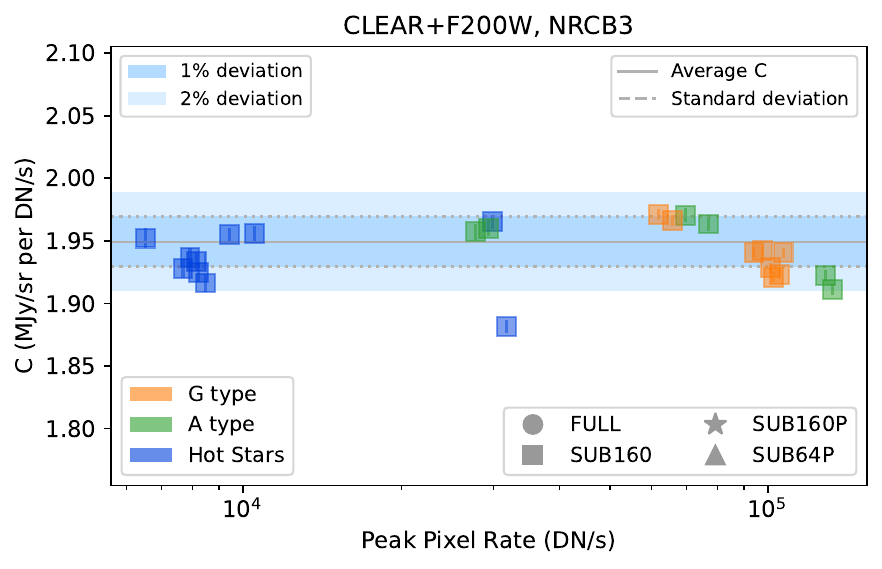}
    \caption{The F200W calibration factor for NRCB3, as a function of the peak pixel rate. The points are measurements from each image with colors indicating the stellar type (solar, A dwarf, or hot stars) and the shapes indicating the subarray. Gray points are targets that were not included in the average. The gray solid line marks the average calibration factor ($C$) and the dashed lines mark the standard deviation. Detector offsets discussed in \S\ref{sec:detoff} are not included here. Note that the x-axis is set to log or linear, as appropriate for each panel. The complete figure set (42 images, all filters/detectors) is available in the online journal.\label{fig:peak}}
\end{figure}

\begin{figure}
    \centering
    \includegraphics[width=\columnwidth]{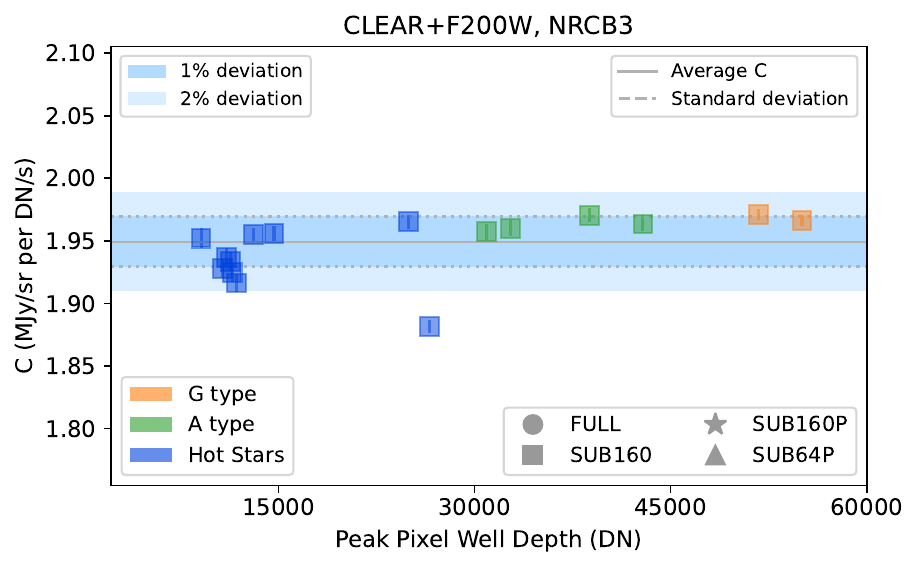}
    \caption{Same as Figure~\ref{fig:peak}, but for the peak pixel well depth. The complete figure set (42 images, all filters/detectors) is available in the online journal. The maximum x-axis value is set to the approximate saturation level (no measurements are saturated). \label{fig:well}}
\end{figure}

\subsubsection{Detector Properties}
\label{sec:det}

Figures~\ref{fig:peak} and \ref{fig:well} show how the calibration factor varies with the peak pixel rate (DN/s) and the pixel well depth (DN) for F200W on NRCB3 (all filters and detectors are included in the figure set). In these plots, each point represents a measurement on an individual image. Colors represent the stellar type and shapes represent the subarray.  The peak pixel rate is a straightforward measurement of the pixel value at the peak of the point source. The well depth was derived by multiplying the peak pixel rate by the length of each integration. NIRCam integrations are sampled along a ``ramp"\footnote{See JDox page on \href{https://jwst-docs.stsci.edu/jwst-near-infrared-camera/nircam-instrumentation/nircam-detector-overview/nircam-detector-readout-patterns\#gsc.tab=0}{NIRCam Detector Readout Patterns}} and some observations saturate before the end of a ramp. For these cases, we recovered the true integration time by examining the \texttt{*\_ramp.fits} files to determine how many groups were observed before saturation. These files can be saved by setting the save\_calibrated\_ramp option to True in the Stage-1 pipeline. 

For all filter+detector combinations, no trends are apparent in either the peak pixel rate or the well depth.

\begin{deluxetable*}{ccll}[h!]
\tablecaption{Calibration Factors Delivered in pmap 1490 \label{tab:calfacs}}
\tablehead{
   \colhead{Row} & \colhead{Units} & \colhead{Label} & \colhead{Description}
}
\startdata
1 & \nodata & Pupil+Filter & The pupil and filter elements \\
2 & \nodata & Mask & The coronagraphic mask. 'None' if not applicable. \\ 
2 & \nodata & Detector & Detector name\\
3 & \nodata & Mode & Observing Mode (Img, Cor, or TAQ) \\
4 & \nodata & Subarray & Subarray name\\
5 & MJy/sr per DN/s & PHOTMJSR & Calibration factor ($C$) for the indicated subarray\\
6 & MJy/sr per DN/s & PHOTMJSR\_ERR & Measurement uncertainty in $C$ for the indicated subarray\\
7 & MJy/sr per DN/s & PHOTMJSR\_STD & Standard deviation for $C$ for the indicated subarray.\\
8 & MJy/sr per DN/s & PHOTMJSR\_SEM & Standard error of the mean for $C$ for the indicated subarray.\\
9 & Magnitude & ZP\_Vega-Sirius & Zeropoint magnitude in the Vega-SIRIUS system\\
10 & Jy & Vega-Sirius\_Jy & Flux density of Sirius from CALSPEC2 model sirius\_stis\_005.fits, scaled to Vega\\
11 & Magnitude & ZP\_Vega & Zeropoint magnitude in the Vega system\\
12 & Jy & Vega\_Jy & Flux density of Vega from CALSPEC2 model alpha\_lyr\_stis\_011.fits\\
13 & Magnitude & ZP\_AB & Zeropoint magnitude in the AB system\\
14 & Steradians & PIXAR\_SR & Average pixel area in steradians\\
15 & \nodata & N\_stars & Number of standard stars included in PHOTMJSR average \\
\enddata
\tablecomments{The calibration factors for subarray observations include the correction factors derived in \S\ref{sec:fullsub}. Mask is blank for non-coronagraphic imaging. Mode is ``Img" for Imaging (including Time Series), ``Cor" for Coronagraphy, and ``TAQ" for Target Acquisition.  PHOTMJSR\_STD and PHOTMJSR\_SEM are both zero when fewer than 3 stars are included in the average. NIRCam filters can be found on \href{https://jwst-docs.stsci.edu/jwst-near-infrared-camera/nircam-instrumentation/nircam-filters\#gsc.tab=0}{JDox}. This table is published in its entirety in the electronic edition of the journal.  A portion is shown here for guidance regarding its form and content. }
\end{deluxetable*}

\begin{figure}
    \includegraphics[width=\columnwidth]{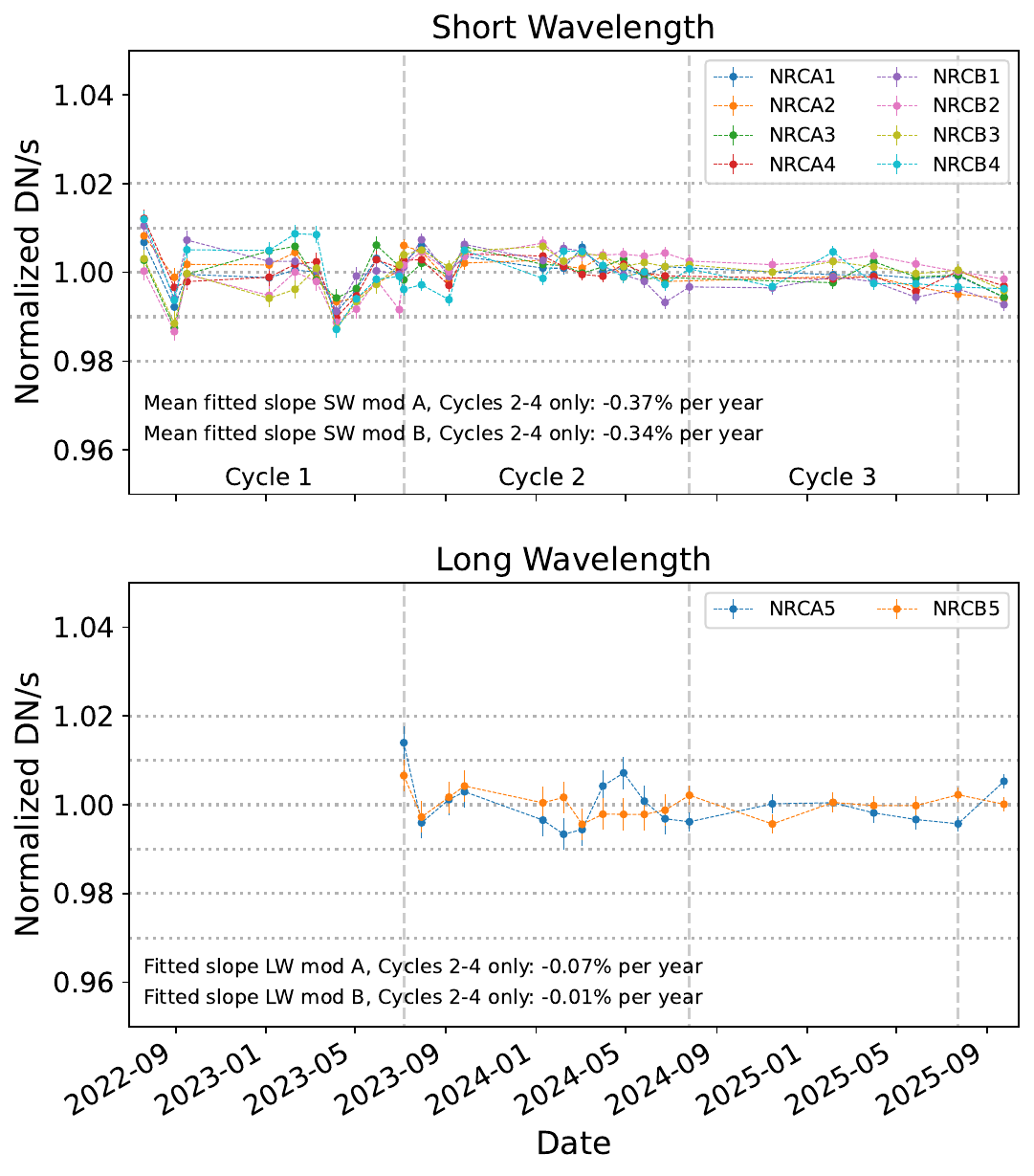}
    \caption{Repeatability of each detector. 
    The lower left indicates the slope measured over Cycles 2--4. Cycle~1 is excluded in the SW channel slope because insufficient dithering caused noisy data. Cycle~1 is also excluded from the slope for the LW channel because the Cycle~1 data were taken in SUB160, and the lack of reference pixels caused noisy data. We switched to SUB160P in Cycle 2.}
    \label{fig:trending}
\end{figure}

\subsection{Repeatability}
\label{sec:time}

NIRCam's responsivity is expected to decline over the lifetime of JWST due primarily to degradation of the mirrors through micrometeoroid impacts and also possibly from degradation of the detectors over time. We tested the repeatability by regularly measuring the same star on each detector since the beginning of science operations.

Figure~\ref{fig:trending} shows the results of the repeatability observations described in \S\ref{sec:repeatdata} from the end of Commissioning (2022 Jul) through 2025 Sep.  Each point is the sigma-clipped average photometry measured on individual dithers, normalized to the average of all epochs. The SW data in Cycle 1 is noisy due to insufficient dithering, so we measure the slope only starting in Cycle~2.  We find a count rate decrease of $\lesssim$0.4\% per year for both modules.  All eight SW detectors follow the same trend. 

In the LW channel, the lack of reference pixels on SUB160 caused noisy data in Cycle 1.  In Cycle~2, we switched to the SUB160P subarrays.  From that point, the count rate has decreased very little, $<$0.1\% per year on both modules. Both the SW and LW count-rate decreases are within the calibration factor uncertainties (see Fig.~\ref{fig:sem}), so 
we do not yet incorporate this count-rate loss in the reference file delivery.  Future deliveries will include this time dependency.

\section{CRDS Reference File Delivery}
\label{sec:crds}

The flux calibration described in this paper was delivered to CRDS on 2026 March 16, as part of pmap 1490. These calibration factors are applied in the \texttt{photom} step in Stage-2 of the pipeline and stored in the fits headers as keyword PHOTMJSR. Previous deliveries included only a single calibration factor for each filter+detector combination -- for this delivery, we include subarray-dependent values using the subarray offsets described in \S\ref{sec:fullsub}. These are applied automatically by the pipeline depending on which subarray is used for an observation.  The delivered calibration factors also include the detector offsets described in \S\ref{sec:detoff}, but they do {\em not} include the time dependence discussed in \S\ref{sec:time}. 

\subsection{Delivery Notes/Caveats}
\label{sec:crds_notes}

\subsubsection{Missing Calibration Factors}

Every allowable combination of detector, filter, mask, and weak lens is included in this delivery, with the exception of WLP8$+$F150W2 for time series imaging and SW filters paired with the LWB mask for coronagraphy.   Future Cycle~5 calibration observations and the associated CRDS deliveries will include measurements of both. In the meantime, we deliver the calibration factor for WLP8$+$F150W2 that was derived from ground testing. For the LWB mask, we substitute the SW filter $+$ SWB calibration factors for the SW filter $+$ LWB combinations.  A comparison between {\em LW filters} paired with SWB to those paired with the LWB, showed most LW filter calibration factors differed by $<$1\%, with the worst (F300M) differing by 3.2\%. If the SW filters behave similarly, we can expect that the substituted calibration factors are within a few percent of the true values.

\subsubsection{Combined values for Coronagraphic Masks}

We note that, at the time of submission, the pipeline (v1.20.2) can not utilize different calibration factors for different coronagraph masks.  The selection of calibration factor for coronagraphy is driven by the pupil wheel element (MASKRND for round masks or MASKBAR for bar masks), the filter, the detector, and the subarray.  The delivered calibration factors for the subarrays {\em are} mask dependent because their locations are associated with specific masks (Fig.~\ref{fig:coron}), so using a subarray ensures that the correct calibration factor is used by the pipeline. However, if using the FULL frame, the pipeline can only ingest one calibration factor for a given pupil+filter+detector combination, regardless of the mask. For example, observations using MASKRND+F356W+NRCALONG on the FULL frame will have only one calibration factor for all three round masks: 210R, 335R, and 430R. For the FULL frame, we therefore delivered calibration factors that are {\em averages} of the round or bar masks.  

\begin{figure}
    \includegraphics[width=\columnwidth]{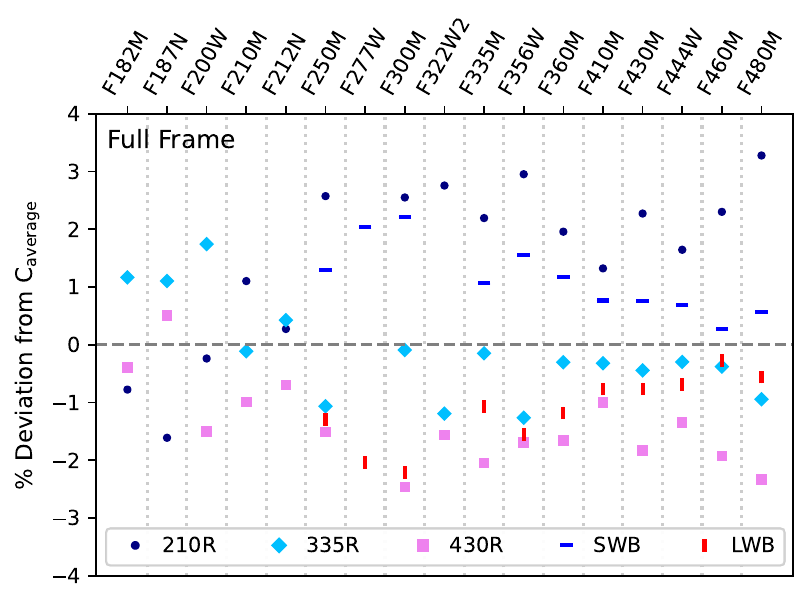}
    \caption{The percent deviation from the average full-frame calibration factor for each mask. Individual calibration factors are within $\sim$3\% of the average value delivered to CRDS for observations in the FULL frame. \label{fig:coron_avg}}
\end{figure}

\begin{figure}
    \includegraphics[width=\columnwidth]{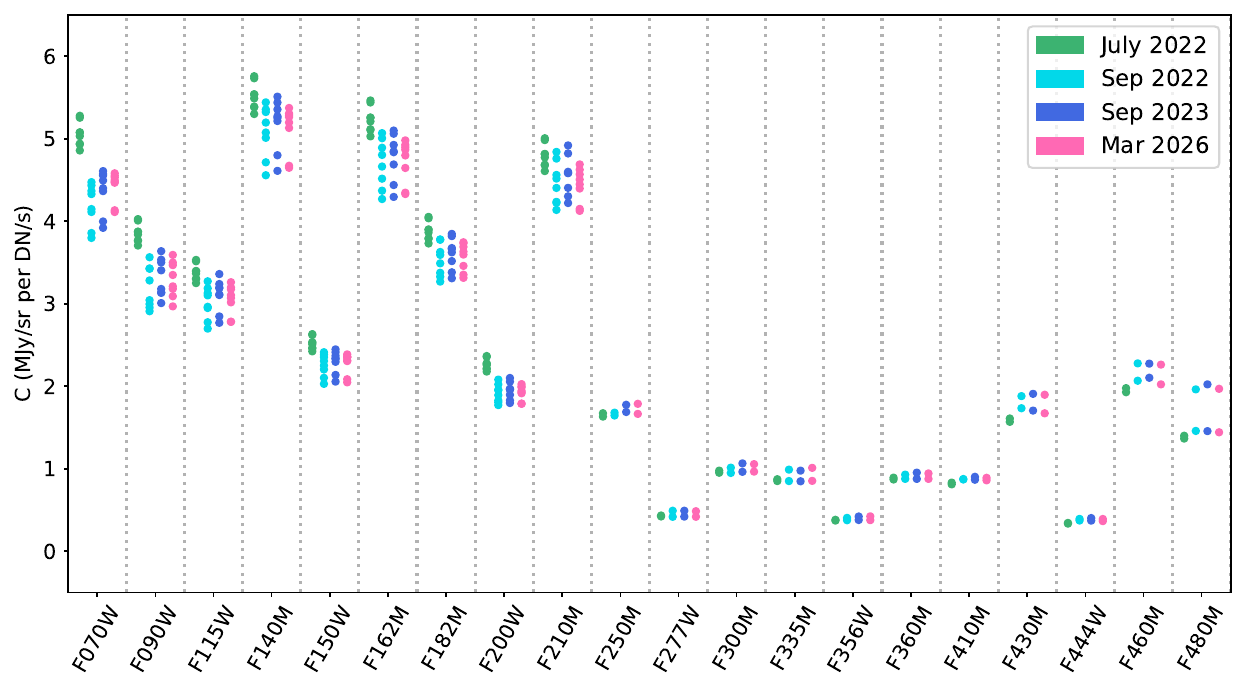}
    \caption{Differences between each CRDS \texttt{photom} reference file delivery for imaging (FULL frame). The dots are the PHOTMJSR values measured on each detector (8 SW, 2 LW detectors). The biggest change occurred in the 2022 Sep delivery. Subsequent deliveries have been relatively stable.  Figures~\ref{fig:diff} and \ref{fig:corondiff} show the percent differences between the 2023 Sep delivery and the new delivery. \label{fig:crds}}
\end{figure}

In Figure~\ref{fig:coron_avg}, we plot the delivered averages compared to the individual mask values and show that the averages are within about 3\% of the individual values.  All mask-dependent calibration factors are published in this paper (Table~\ref{tab:calfacs}) and on JDox\footnote{\href{https://jwst-docs.stsci.edu/jwst-near-infrared-camera/nircam-performance/nircam-absolute-flux-calibration-and-zeropoints\#gsc.tab=0}{JDox page} that lists NIRCam calibration factors} so that users can correct their images using the appropriate calibration factors by dividing calibrated images by the PHOTMJSR keyword in the header and multiplying by the values reported here before proceeding with Stage~3 of the pipeline. Future versions of the pipeline will incorporate the ability to select the calibration factor based on the mask. At that point, we will deliver new reference files with the individual mask values for the FULL frame.

\begin{figure}
    \begin{subfigure}{\columnwidth}
    \centering
        \includegraphics[width=\columnwidth]{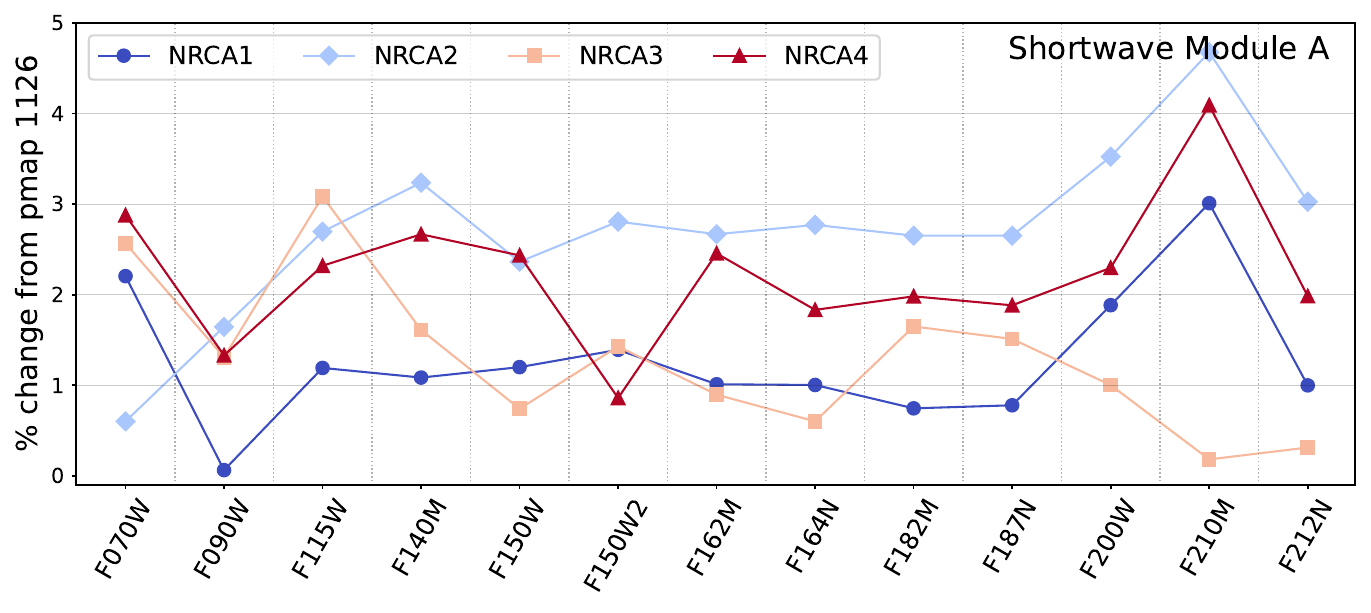}
    \end{subfigure}   
    \begin{subfigure}{\columnwidth}
    \centering
        \includegraphics[width=\columnwidth]{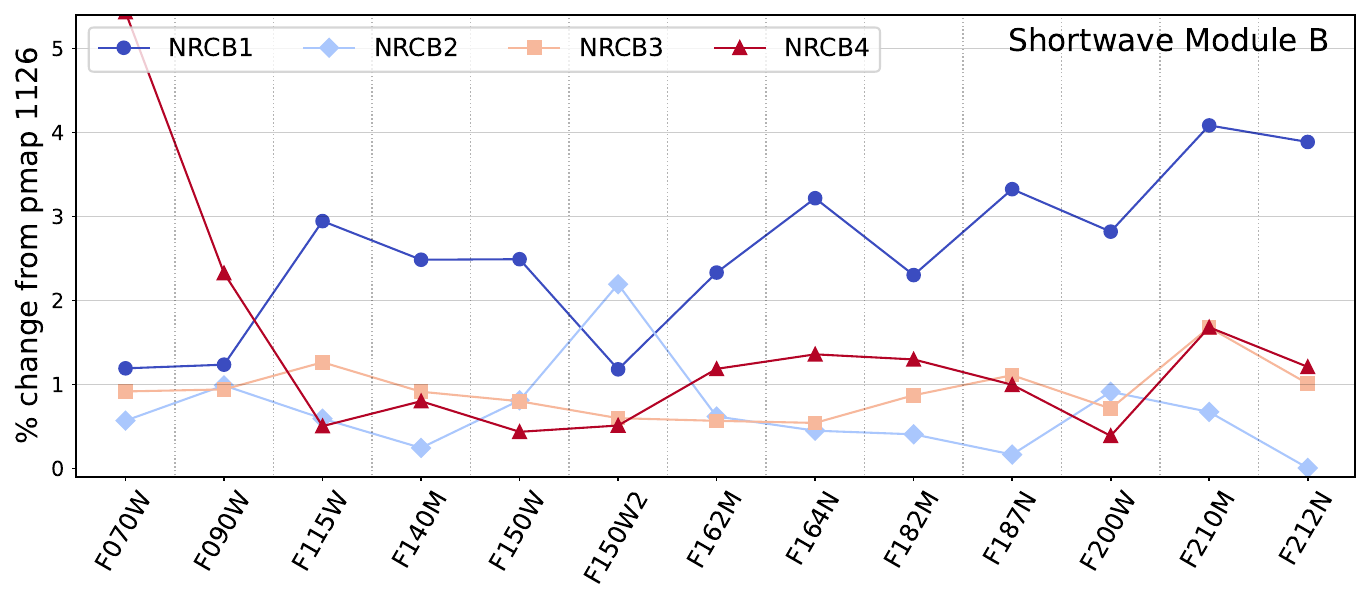}
    \end{subfigure}   
    \begin{subfigure}{\columnwidth}
    \centering
        \includegraphics[width=\columnwidth]{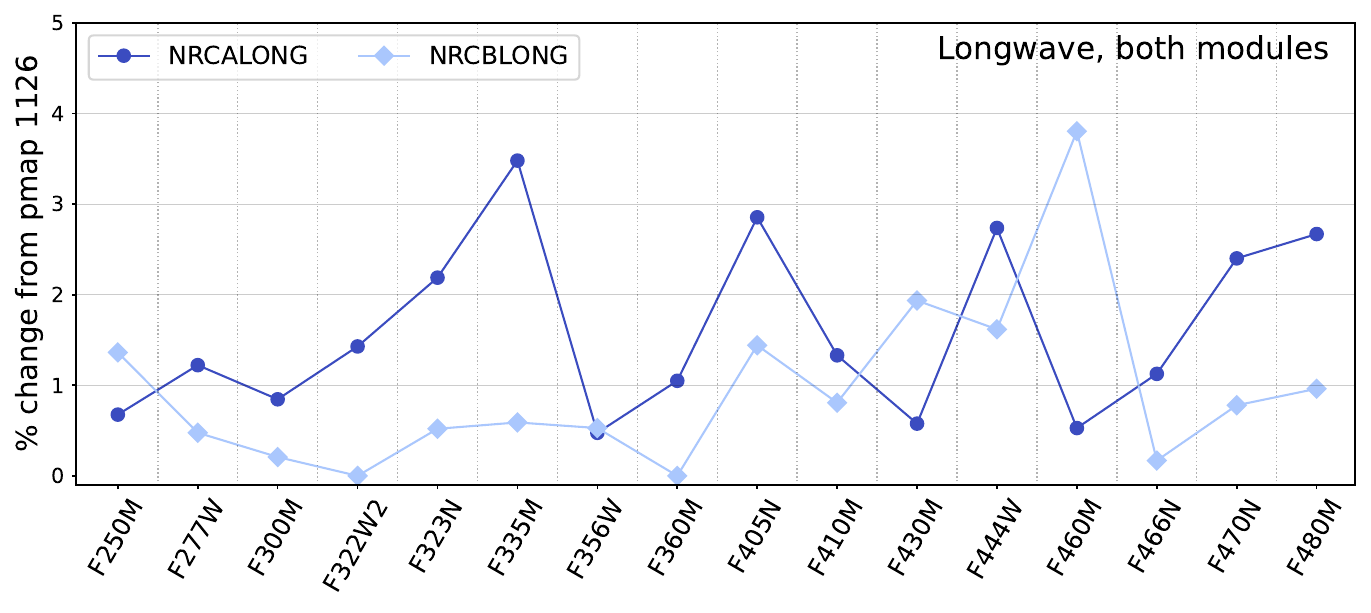}
    \end{subfigure}  
    \caption{The absolute value of the percent differences between the FULL frame calibration factors for this delivery (pmap 1490) and the previous delivery (pmap 1126) for imaging filters. Most calibration factors changed by $<$3\%. \label{fig:diff}}
\end{figure}

\begin{figure}
    \begin{subfigure}{\columnwidth}
    \centering
        \includegraphics[width=\columnwidth]{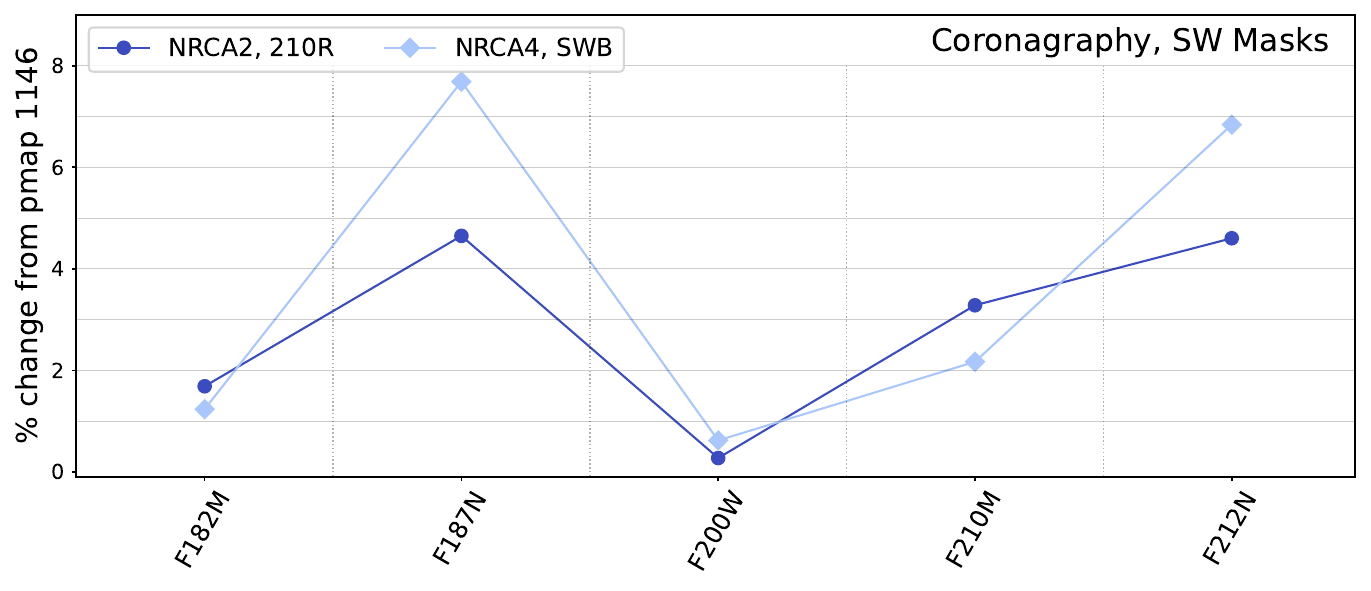}
    \end{subfigure}   
    \begin{subfigure}{\columnwidth}
    \centering
        \includegraphics[width=\columnwidth]{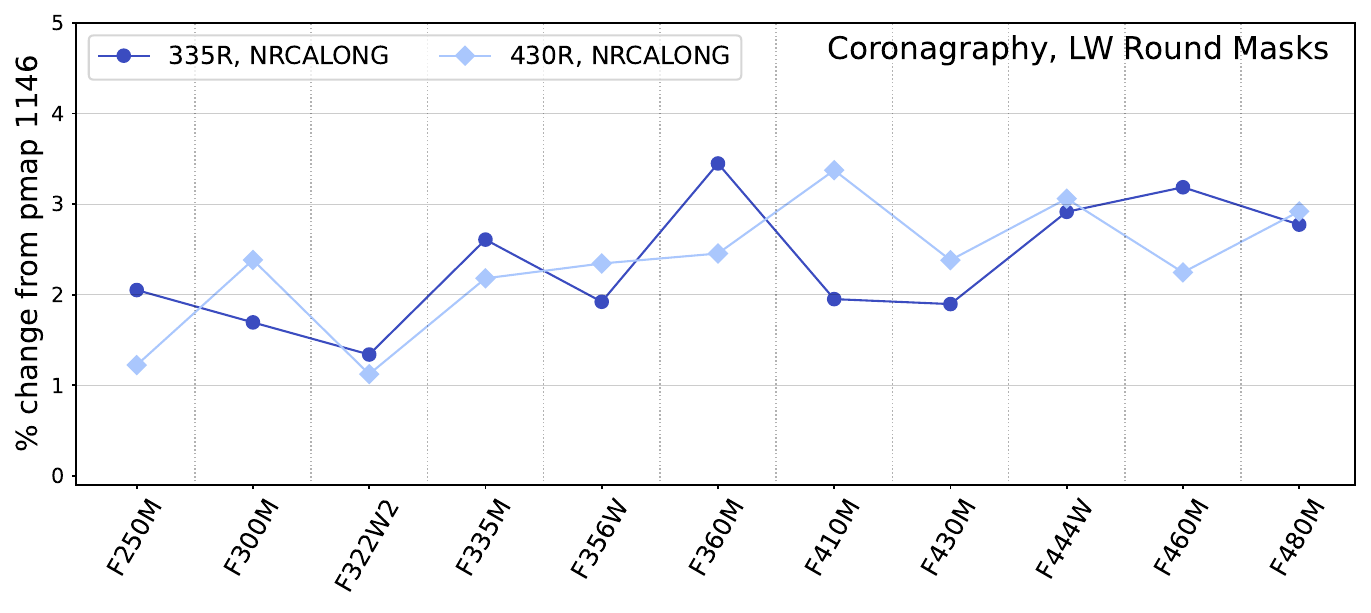}
    \end{subfigure}   
    \begin{subfigure}{\columnwidth}
    \centering
        \includegraphics[width=\columnwidth]{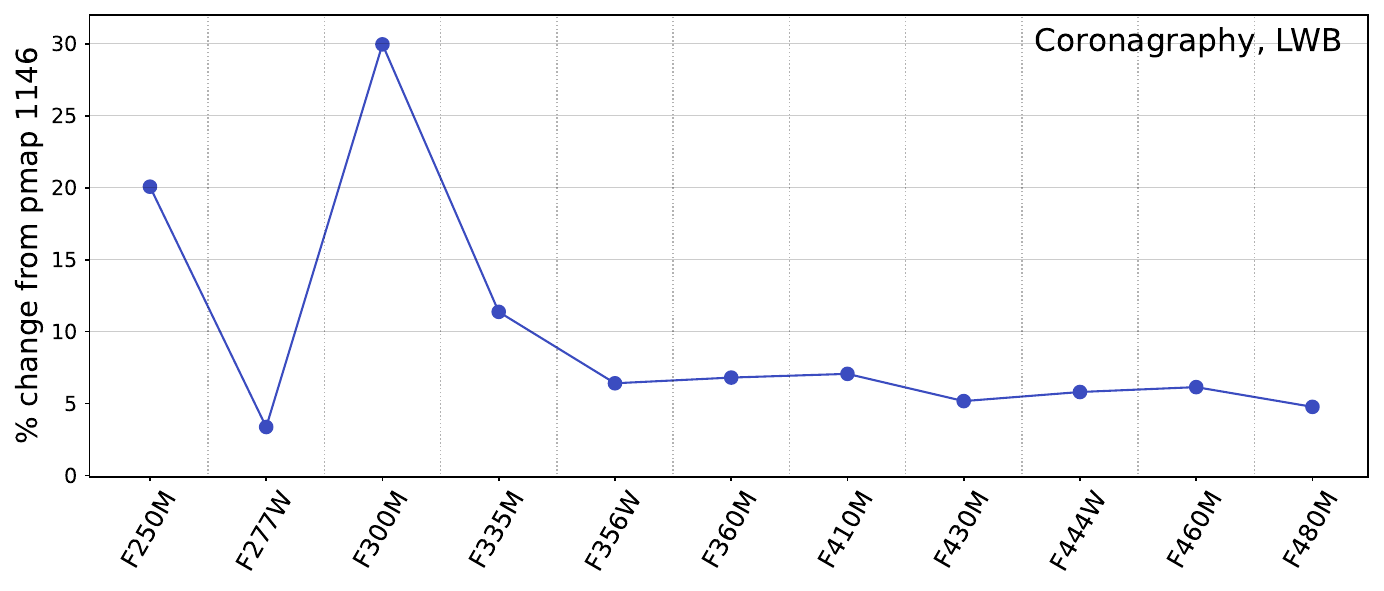}
    \end{subfigure}  
    \caption{The absolute value of the percent differences in the FULL frame calibration factors between this delivery (pmap 1490) and the previous delivery for coronaraphy (pmap 1146). Note that the previous delivery did not include dual-channel coronagraphy data, so this plot includes only comparisons in the primary coronagraph channel (SW filters with SW masks or LW filters with LW masks). \label{fig:corondiff}}
\end{figure}

\subsection{Comparison to Previous Deliveries}
\label{sec:diff}

In Figures \ref{fig:crds}-\ref{fig:corondiff}, we show comparisons between previous CRDS deliveries and the delivery described in this paper.  Figure \ref{fig:crds} shows how the values have changed since Commissioning and shows that while there was a sizeable change between the imaging calibration factors derived in July and September of 2022, values have remained relatively stable since then.  

Figure~\ref{fig:diff} shows the absolute value of the percent difference between the previous delivery and the delivery described here.  For imaging, differences are $<$3\% for most filters/detectors, with some showing changes up to 4\%--5\%. The largest change is for F070W on NRCB4 ($\sim$5.5\%), with this large shift due in part to the applied detector offsets described in \S\ref{sec:detoff}. 

For coronagraphy (Fig.~\ref{fig:corondiff}), the differences are $<$3\%--4\% for the 335R and 430R masks in the LW channel and $<$5\% for the 210R mask in the SW channel. The differences are larger for the bar masks because the Cycle 1 targets were partially obscured by those masks (see \S\ref{sec:coron_results}). The affected data have been excluded in the delivery described here, and thus the differences in the calibration factors are high in some filters. For the SWB, the differences are as high as $\approx$8\% for the narrow filters, but are $<$2\% for the other filters. For the LWB, most filters show differences of $<$5\%, though the bluest filters show differences as high as 30\%. While the LWB mask is narrower on the blue end, the star was actually {\em more} occulted in the blue filters because the Cycle 1 target offsets placed the star almost exactly on the first minimum of the bar transmission.\footnote{See Figure~2 on the \href{https://jwst-docs.stsci.edu/jwst-near-infrared-camera/nircam-observing-modes/nircam-coronagraphic-imaging\#gsc.tab=0}{NIRCam Coronagraphic Imaging JDox page}.}  For the redder filters, the same offset placed the star just inside this minimum, where transmission is higher.

Similar comparisons do not exist for the weak lens elements or for dual-channel coronagraphy (filters paired with masks in the opposite channel) because this is the first on-sky delivery of calibration factors for both.

\subsection{Magnitude Zeropoints}

The JWST pipeline employs ``Vega" zeropoints that use Sirius as the color reference \citep{Rieke+2022, Gordon+2022}. These zeropoints are stored in the \texttt{abvegaoffset} reference file and are used to compute Vega-Sirius magnitudes for the catalog output in Stage-3 of the pipeline.  We list these zeropoints along with AB~mag zeropoints in Table~\ref{tab:calfacs}, along with the traditional Vega zeropoints for convenience. The Vega-Sirius zeropoints are tied to the \texttt{sirius\_stis\_005.fits} spectrum in the CALSPEC2 database, while the traditional Vega zeropoints are tied to the \texttt{alpha\_lyr\_stis\_011.fits} spectrum. To compute the zeropoints ($Z$) for the \texttt{abvegaoffset} reference file, we follow:

\begin{equation}
    Z_{\rm Vega-Sirius} = -2.5\, {\rm log}_{10}\frac{C}{(f_\nu^{\rm Sirius} / \Omega_{\rm pix}) - 1.395},  
\end{equation}

\noindent where $f_\nu$ is the flux density of Sirius in MJy, $C$ is the calibration factor in MJy/sr per DN/s (PHOTMJSR in the image headers), $\Omega_{\rm pix}$ is the average solid angle per pixel in steradians (PIXAR\_SR in the image headers), and $-1.395$ is the K$_{\rm s}$ magnitude of Sirius from \citet{Rieke+2022}. The Sirius CALSPEC2 model defines magnitude $-1.395$ at all wavelengths, so $m_{\rm Vega-Sirius} = m_{\rm Sirius} - 1.395$. The Vega-Sirius magnitude is then defined as:

\begin{equation}
m_{\rm Vega-Sirius} = m_{\rm inst} + Z_{\rm Vega-Sirius},
\end{equation}

\noindent where m$_{\rm inst}$ is the instrumental magnitude, defined as

\begin{equation}
m_{\rm inst} = -2.5\, {\rm log}_{10}\,f_{\rm inst}.
\end{equation}

\noindent Here, $f_{\rm inst}$ is the instrumental flux in DN/s. Alternatively, these equations can be reorganized to compute the Vega-Sirius magnitude directly from the measured flux ($f_{\rm measured}$, in MJy/sr):

\begin{equation}
    m_{\rm Vega-Sirius} = -2.5\, {\rm log}_{10}\frac{f_{\rm measured}\,\Omega_{\rm pix}}{10^{-6}\,f_\nu^{Vega-Sirius}},
\end{equation}

\noindent where  $f_\nu^{\rm Vega-Sirius}$ is the flux density of Sirius in Jy, scaled to Vega ($f_\nu^{\rm Vega-Sirius} = f_\nu^{\rm Sirius}/3.614$), which is listed in Table~\ref{tab:calfacs} as Vega-Sirius\_Fnu (column 10).

\section{Summary}

We present the latest absolute flux calibration for the JWST/NIRCam Imaging, Time Series Imaging and Coronagraphy modes. This analysis includes observations of photometric standard stars spanning 3.5 years, from Cycles 1--4 (2022--2025), taken as part of the JWST Absolute Flux Calibration Program described in \citet{Gordon+2022}. We include three types of standard stars: 9 hot stars, 5 solar analogs, and 5 A~dwarf stars.  This analysis includes all 13 SW filters on the 8 SW detectors and all 16 LW filters on the 2 LW detectors. We also include both weak lenses, which can be used in the TS imaging modes.  For coronagraphy, we include all 5 coronagraphic masks, each paired with every filter allowed for that mode, with the exception of the LWB paired with SW filters.  

We derived calibration factors by comparing the predicted stellar fluxes from CALSPEC2 models to the measured fluxes derived from aperture photometry, using custom aperture corrections determined from STPSF models. 
Automated routines that produce the aperture photometry, the STPSF models, and the calibration factors are available online.\footnote{\href{https://github.com/spacetelescope/nircam-fluxcal}{https://github.com/spacetelescope/nircam-fluxcal}}

In summary, we find that:
\vspace{-0.5em}
\begin{itemize}[itemsep=2pt, parsep=0pt, leftmargin=2em]
    \item the NIRCam detectors show a small decrease in the count-rate of $\lesssim$0.4\% per year in the SW channel and $<$0.1\% per year in the LW channel, 
    \item the scatter in the calibration factors is $<$2\% for most of the imaging filters, and $<$1\% for about half of them,
    \item For weak lenses and coronagraphy, the scatter is $<$1.5\% for most setups that have measurements in at least 3 stars, and
    \item for most filters/masks, the calibration factors changed by $<$4\% from the previous delivery. 
\end{itemize}

The delivered calibration factors include corrections for offsets between subarrays and the full frame, up to $\approx$1\%.  For 8 filters (4 SW and 4 LW), we also applied offsets between detectors measured from LMC and 47\,Tuc data, on the order of $\sim$1--2\%. 
A Cycle 5 calibration program will measure offsets for the remaining filters. We note that the final calibration factors do {\em not} include the slight decrease in the count-rate, since the overall decrease is still within the uncertainties of the calibration factors.

For the standard imaging filters, 
some of the scatter in the calibraiton factors appears to be caused by uncertainties in the flat field. We find no dependence on the stellar type or on detector effects such as well depth. 

For the weak lenses and coronagraphic masks, only about a third of the filter+detector[+mask] combinations have measurements of 3 or more stars. This is due to issues with the observations and/or to the fact that dual-channel coronagraphy was not enabled until Cycle 2. 
Future Cycle 5 calibration programs will focus on obtaining measurements of additional standard stars for all weak lens and coronagraphy settings.

The calibration factors described here were delivered to CRDS on 2026 Mar 16 as part of pmap 1490. For most filter+detector combinations, the change from the previous delivery is $<$4\%. The exceptions are: F070W on NRCB4, which changed by $\approx$5.5\%, two measurements in the SWB mask that show a 7\%--8\% change, and all LWB mask measurements, which show changes as high as 30\%. The large changes for the SWB and LWB masks are due to the removal of measurements where the bar masks appear to partially obscure the star, which had been included in the previous delivery.

The NIRCam flux calibration will continue to improve as we collect additional observations of flux standard stars in upcoming cycles and as we make improvements to related calibrations. Future deliveries will include additional analysis of the flat fields, an assessment of detector offsets in all filters, and measurements of more standard stars.

\bigskip

\begin{acknowledgements}
 We thank Karl Gordon, Kevin Volk, and the rest of the JWST Absolute Flux Coordination Team for extremely useful discussions and for helping to design the observing program described in \citep{Gordon+2022}. We thank the referee for helpful comments. This work is based on observations made with the NASA/ESA/CSA James Webb Space Telescope. The data were obtained from the Mikulski Archive for Space Telescopes at the Space Telescope Science Institute, which is operated by the Association of Universities for Research in Astronomy, Inc., under NASA contract NAS 5-03127 for {\it JWST}. These observations are associated with programs 1069, 1476, 1536, 1537, 1538, 1539, 4452, 4496, 4497, 4498, 4499, 6605, 6606, 6607, 6630, 6631, 7487, 7615, 7671, and 8882. The data can be accessed at  
\dataset[doi:10.17909/0evw-me44]{https://doi.org/10.17909/0evw-me44},
\dataset[doi:10.17909/xa3a-fr32]{https://doi.org/10.17909/xa3a-fr32},
\dataset[doi:10.17909/apw1-5341]{https://doi.org/10.17909/apw1-5341},
\dataset[doi:10.17909/q222-ak34]{https://doi.org/10.17909/q222-ak34}, \dataset[doi:10.17909/q0t9-d777]{https://doi.org/10.17909/q0t9-d777}, and \dataset[doi:10.17909/dkpn-h062]{https://doi.org/10.17909/dkpn-h062}.
\end{acknowledgements}

\vspace{5mm}
\facilities{JWST (NIRCam), MAST}

\software{NIRCam fluxcal \citep{scripts}, astropy \citep{astropy1, astropy2, astropy3},
         scikit-learn \citep{scikit-learn}, scipy \citep{SciPy2020}, photutils \citep{Bradley+2023}, STPSF \citep{webbpsf}, Synphot \citep{synphot}, JWST calibration pipeline \citep{pipeline}, hst1pass \citep{Anderson+2022} and One-Pass-Fitting \citep{Bajaj_One-Pass} }

\newpage



\appendix
\restartappendixnumbering

\section{Observation Details}

Table~\ref{tab:targs} summarizes the targets that were observed for each mode, including information about the filters, subarrays, and detectors.  Tables~\ref{tab:1536}--\ref{tab:8882} summarize each of the observing programs in more detail.  

\begin{deluxetable*}{lllllll}[h!]
\tabletypesize{\tiny}
\tablecaption{Summary of targets, modes, filters, and detectectors\label{tab:targs}}
\tablehead{\colhead{Mode} & \colhead{Subarrays} & \colhead{Elements} & \colhead{Detectors} & \colhead{A dwarfs} & \colhead{Hot Stars} & \colhead{Solar}}
\startdata
\multicolumn{7}{c}{----- Part 1 -----}\\
Imaging, TS & SUB160, SUB160P, SUB64P & All SW+LW filters & All detectors &	J1743045 & GD\,71 & P330E  \\	
TS &SUB400P, SUBGRISM256, FULL& WLP4/8 +SW filters\tablenotemark{a} & NRCA1, NRCA3, NRCB1	& J1743045	& G\,191-B2B & P330E	\\
Coronagraphy & SUB640, FULL & 210R +SW filters &  NRCA2	& J1743045	& G\,191-B2B & P330E  \\	
Coronagraphy & SUB640, FULL & 210R +LW filters & NRCALONG	& J1743045	& \nodata & P330E  \\	
Coronagraphy & SUB320 & 335R +SW filters & NRCA2	& \nodata &	\nodata & P330E  \\	
Coronagraphy & SUB320 & 335R +LW filters &  NRCALONG	& J1743045	& G\,191-B2B & P330E  \\
Coronagraphy & SUB320 & 430R +SW filters & NRCA2	& \nodata	& \nodata	& P330E \\	
Coronagraphy & SUB320 & 430R +LW filters & NRCALONG	& J1743045	& G\,191-B2B & P330E  \\	
Coronagraphy & SUB640, FULL & SWB +SW filters &  NRCA4	& J1743045	& G\,191-B2B\tablenotemark{b} &	P330E   \\
Coronagraphy & SUB640, FULL & SWB +LW filters & NRCALONG	& \nodata & \nodata & P330E \\	
Coronagraphy & SUB400X256 & LWB +SW filters & NRCA4	& \nodata &	\nodata & P330E\tablenotemark{c}	 \\
Coronagraphy & SUB400X256 & LWB +LW filters & NRCALONG	& J1743045\tablenotemark{d} & G\,191-B2B\tablenotemark{d} & P330E	 \\
Coronagraph TA & FS subarrays & F210M/F335M & NRCA2, NRCA4, NRCALONG& J1743045 & G\, 191-B2B & P330E  \\	
Coronagraph TA & ND subarrays & F210M/F335M & NRCA2, NRCA4, NRCALONG& HR 5467 & 10 Lac & HR 6538	 \\
\hline
\multicolumn{7}{c}{----- Part 2 -----}\\
Imaging & SUB160 & subset of M and W filters & NRCA3+NRCALONG & J1757132 & G\,191-B2B & C26202  \\
 & SUB160P && NRCB1+NRCBLONG & J1802271 & GD\,153  & P177D  \\
 & SUB64P &&& J1805292 & LDS 749B & SNAP-2  \\
 & FULL &&& & WD1057 &  \\
 && && & WDFS0122 &  \\
 && && & WDFS0458 &  \\
 && && & WDFS2317 &  \\
\enddata
\tablenotetext{a}{\ Except WLP8+F150W2 on NRCB1.}
\tablenotetext{b}{\ G\,191-B2B excluded from analysis (too close to the SWB mask).}
\tablenotetext{c}{\ P330E excluded from analysis (too close to the subarray edge).}
\tablenotetext{d}{\ G\,191-B2B and J1743045 excluded from analysis (too close to the LWB mask).}
\tablecomments{\ For Part 1, all allowed filters and detectors were observed, unless otherwise noted. For Part 2 of the program, the filters and subarrays varied for each target. Time Series (TS) includes the Grism Time Series and Imaging Time Series modes.  The Grism TS mode includes an option to perform imaging in the SW channel with the WLP8 and WLP4 weak lenses on NRCA1 and NRCA3. The Imaging TS mode allows imaging with WLP8 on NRCB1, along with imaging in all of the standard imaging filters on NRCB1/NRCBLONG.}
\end{deluxetable*}

\begin{deluxetable*}{lrccccl}[h!]
\tabletypesize{\footnotesize}
\tablecaption{Cycle 1 Observations of A Dwarf Stars, PID 1536 \label{tab:1536}}
\tablehead{
\colhead{Target} & \colhead{Obs\#} & \colhead{Date} & \colhead{Detector(s)} & \colhead{Subarray} & \colhead{Pupil\tablenotemark{a}} & \colhead{Filters}
}
\startdata
\multicolumn{7}{c}{----- Part 1 Imaging -----}\\
J1743045 & 150 & 2022 Sep 17 & B1--B4 & SUB160 & CLEAR &  All SW except F150W2\\
         &     &             & BLONG     & SUB160 & CLEAR &  All LW except F322W2\\
J1743045 & 248 & 2022 Oct 19 & A1--A4 & SUB160 & CLEAR & All SW except F150W2 \\
         &     &             & ALONG & SUB160 & CLEAR & All LW except F322W2 \\
J1743045 & 49 &2022 Aug 16& B1    & SUB64P & CLEAR & F070W, F150W2 \\
         &    &           & BLONG & SUB64P & CLEAR & F356W, F322W2 \\
J1743045 & 51 &2022 Sep 03& A3    & SUB64P & CLEAR & F070W, F150W2\\
         &    &           & ALONG & SUB64P & CLEAR & F277W, F322W2 \\
\hline
\multicolumn{7}{c}{----- Part 2 Imaging -----}\\
J1802271 & 60 &2022 Jul 08 & B1--B4 & SUB160 & CLEAR & F070W, F200W, F140M, F162M, F182M \\
         &    &            & BLONG  & SUB160 & CLEAR & F277W, F356W, F444W, F250M, F460M \\
J1757132 & 61 & 2022 Sep 03 & B1--B4 & SUB160 & CLEAR & F070W, F200W, F140M, F162M, F182M \\
         &   &              & BLONG  & SUB160 & CLEAR & F277W, F356W, F444W, F250M, F460M \\
J1757132 & 78 & 2022 Sep 02 & B1    & SUB64P & CLEAR & F070W, F140M \\
         &    &             & BLONG & SUB64P & CLEAR & F277W, F356W \\
\hline
\multicolumn{7}{c}{----- Part 1 Weak Lens Imaging -----}\\
J1743045  & 80 &2002 Aug 20& A1, A3& SUB320 & CLEAR & WLP4$+$F212N2 \\
          &    &           & A1, A3& SUB320 & WLP8 & F070W, F140M, F182M, F210M, F187N, F212N \\
%
J1743045  & 79 &2022 Sep 02& B1    & SUB400P & WLP8 & F150W, F200W, F140M, F182M, F210M, F187N, F212N \\
%
\hline
\multicolumn{7}{c}{----- Part 1 Coronagraphic Imaging -----}\\
J1743045  & 63 &2022 Sep 03& A4 & SUB640 & MASKSWB & F200W, F182M, F210M, F187N, F212N \\
J1743045  & 64 &2022 Sep 03& A2 & SUB640 & MASK210R & F200W, F182M, F210M, F187N, F212N \\
J1743045  & 65 &2022 Sep 03& ALONG & SUB320 & MASKLWB & All LW W and M filters \\
J1743045  & 66 &2022 Sep 03& ALONG & SUB320 & MASK335R & All LW W2, W, and M filters except F277W \\
J1743045  & 67 &2022 Sep 03& ALONG & SUB320 & MASK430R & All LW W2, W, and M filters except F277W \\
\enddata
\tablenotetext{a}{When not using WLP8 or a coronagraphic mask, most filters are paired with the CLEAR element in the pupil wheel. The exception is filters that are {\em in} the pupil wheel, which are paired with wide filters in the filter wheel (Pupil$+$Filter: F164M$+$F150W2, F162N$+$F150W2, F323N$+$F322W2, F405N$+$F444W, F460N$+$F444W, and F470N$+$F444W).}
\end{deluxetable*}

\begin{deluxetable*}{lrccccl}[h!]
\tabletypesize{\footnotesize}
\tablecaption{Cycle 1 Observations of Hot Stars, PID 1537 \label{tab:1537}}
\tablehead{\colhead{Target} & \colhead{Obs\#} & \colhead{Date} & \colhead{Detector(s)} & \colhead{Subarray} & \colhead{Pupil\tablenotemark{a}} & \colhead{Filters}}
\startdata
\multicolumn{7}{c}{----- Part 1 Imaging -----}\\
GD 71 & 14 & 2022 Sep 21 & B1--B4 & SUB160 & CLEAR & All SW \\
      &    &             & BLONG  & SUB160 & CLEAR & All LW \\
GD 71 & 15 & 2022 Sep 22 & A1--A4 & SUB160 & CLEAR & All SW \\
      &    &             & ALONG  & SUB160 & CLEAR & All LW \\
\hline
\multicolumn{7}{c}{----- Part 2 Imaging -----}\\
GD 153 & 23 & 2023 Jan 16 & B1--B4 & SUB160 & CLEAR & F070W, F090W, F115W, F150W, F200W \\
       &    &             & BLONG  & SUB160 & CLEAR & F277W, F356W, F444W, F250M, F460M \\
G 191-B2B & 24 & 2022 Sep 08 & B1--B4 & SUB160 & CLEAR & F070W, F090W, F115W, F150W, F200W \\
          &    &             & BLONG  & SUB160 & CLEAR & F277W, F356W, F444W, F250M, F460M \\
\hline
\multicolumn{7}{c}{----- Part 1 Weak Lens Imaging -----}\\
G 191-B2B  & 38 & 2022 Sep 10 & B1     & SUB400P & WLP8   & F150W, F200W, F140M, F182M, F210M, F187N, F212N \\
G 191-B2B  & 39 & 2022 Sep 10 & A1, A3 & SUB320 & CLEAR & WLP4+F212N2 \\
           &    &             & A1, A3 & SUB320 & WLP8    & F070W, F140M, F182M, F210M, F187N, F212N \\
\hline
\multicolumn{7}{c}{----- Part 1 Coronagraphic Imaging -----}\\
G 191-B2B  & 25 & 2022 Sep 08 & A4 & SUB640 & MASKSWB  & F200W, F182M, F210M, F187N, F212N \\
G 191-B2B  & 26 & 2022 Sep 08 & A2 & SUB640 & MASK210R & F200W, F182M, F210M, F187N, F212N \\
G 191-B2B  & 27 & 2022 Sep 09 & ALONG & SUB320 & MASKLWB  & All LW W and M filters \\
G 191-B2B  & 28 & 2022 Sep 09 & ALONG & SUB320 & MASK335R & All LW W2, W, and M filters except F277W \\ 
G 191-B2B  & 29 & 2022 Sep 09 & ALONG & SUB320 & MASK430R & All LW W2, W, and M filters except F277W \\ 
\enddata
\tablenotetext{a}{When not using WLP8 or a coronagraphic mask, most filters are paired with the CLEAR element in the pupil wheel. The exception is filters that are {\em in} the pupil wheel, which are paired with wide filters in the filter wheel (Pupil$+$Filter: F164M$+$F150W2, F162N$+$F150W2, F323N$+$F322W2, F405N$+$F444W, F460N$+$F444W, and F470N$+$F444W).}
\end{deluxetable*}

\begin{deluxetable*}{lrccccl}
\tablewidth{0pt}
\tabletypesize{\footnotesize}
\tablecaption{Cycle 1 Observations of Solar Analog Stars, PID 1538\label{tab:1538}}
\tablehead{\colhead{Target} & \colhead{Obs\#} & \colhead{Date} & \colhead{Detector(s)} & \colhead{Subarray} & \colhead{Pupil\tablenotemark{a}} & \colhead{Filters}}
\startdata
\multicolumn{7}{c}{----- Part 1 Imaging -----}\\
P330-E  & 154  & 2022 Aug 29 & A1--A4 & SUB160 & CLEAR & All SW except F150W2\\
             &      &             & ALONG  & SUB160 & CLEAR & All LW except F322W2\\
P330-E  & 155  & 2022 Aug 29 & B1--B4 & SUB160 & CLEAR & All SW except F150W2\\
             &      &             & BLONG  & SUB160 & CLEAR & All LW except F322W2\\
P330-E  & 56 &2022 Aug 10& B1     & SUB64P & CLEAR & F150W2, F070W \\
             &    &           & BLONG  & SUB64P & CLEAR & F322W2, F356W \\
P330-E  & 57 &2022 Aug 30& A3     & SUB64P & CLEAR & F150W2, F070W \\
             &    &           & ALONG  & SUB64P & CLEAR & F322W2, F356W \\
\hline
\multicolumn{7}{c}{----- Part 2 Imaging -----}\\
P177-D & 53 & 2023 Mar 18 & B1--B4 & SUB160 & CLEAR & F070W, F200W, F140M, F162M, F182M \\
            &    &             & BLONG  & SUB160 & CLEAR & F356W, F444W, F250M, F410M, F460M \\
\hline
\multicolumn{7}{c}{----- Part 1 Weak Lens Imaging -----}\\
P330-E  & 70 &2022 Aug 14 & A1, A3& SUB320 & CLEAR & WLP4+F212N2\\
             &    &            & A1, A3& SUB320 & WLP8  & F070W, F140M, F182M, F210M, F187N, F212N \\
P330-E  & 71 &2022 Aug 14 & B1 & SUB400P & WLP8 & F150W, F200W, F140M, F182M, F210M, F187N, F212N \\
\hline
\multicolumn{7}{c}{----- Part 1 Coronagraphic Imaging -----}\\
P330-E  & 48 &2022 Aug 20& A4 & SUB640 & MASKSWB & F200W, F182M, F210M, F187N, F212N \\
P330-E  & 49 &2022 Aug 20& A2 & SUB640 & MASK210R & F200W, F182M, F210M, F187N, F212N \\
P330-E  & 50 &2022 Aug 20& ALONG & SUB320 & MASKLWB & All LW W and M filters \\
P330-E  & 51 &2022 Aug 21& ALONG & SUB320 & MASK335R & All LW W2, W, and M filters except F277W \\
P330-E  & 52 &2022 Aug 21& ALONG & SUB320 & MASK430R & All LW W2, W, and M filters except F277W \\
\enddata
\tablenotetext{a}{When not using WLP8 or a coronagraphic mask, most filters are paired with the CLEAR element in the pupil wheel. The exception is filters that are {\em in} the pupil wheel, which are paired with wide filters in the filter wheel (Pupil$+$Filter: F164M$+$F150W2, F162N$+$F150W2, F323N$+$F322W2, F405N$+$F444W, F460N$+$F444W, and F470N$+$F444W).}
\end{deluxetable*}

\begin{deluxetable*}{lrccccl}[h!]
\tablewidth{0pt}
\tabletypesize{\footnotesize}
\tablecaption{Cycle 2 Observations of A dwarf Stars, PID 4496\label{tab:4496}}
\tablehead{\colhead{Target} & \colhead{Obs\#} & \colhead{Date} & \colhead{Detector(s)} & \colhead{Subarray} & \colhead{Pupil} & \colhead{Filters}} 
\startdata
\multicolumn{7}{c}{----- Part 2 Imaging -----}\\
J1757132  & 21  & 2024 May 19 & A3     & SUB64P & CLEAR & F070W, F090W, F115W, F150W, F200W \\
          &     &  & ALONG  & SUB64P & CLEAR & F277W, F356W, F444W, F410M, F430M\\
J1805292  & 22  & 2024 Jun 03 & B1     & SUB64P & CLEAR & F070W, F090W, F115W, F150W, F200W \\
          &     &  & BLONG  & SUB64P & CLEAR & F277W, F356W, F444W, F460M, F480M\\
\hline
\multicolumn{7}{c}{----- Part 1 Coronagraphic TA Imaging -----}\\
HR 5467  &  23 & 2024 Mar 28 & A4 & ND Square SWB & MASKSWB & F210M \\
HR 5467  & 124 & 2024 Jun 18 & A4 & ND Square SWBS & MASKSWB & F210M \\
HR 5467  &  25 & 2024 Apr 12 & A2 & ND Square 210R & MASK210R & F210M \\
HR 5467  &  26 & 2024 Apr 24 & ALONG & ND Square LWB & MASKLWB & F335M \\
HR 5467  &  28 & 2024 Jun 18 & ALONG & ND Square 335R & MASK335R & F335M \\
HR 5467  &  29 & 2024 Jun 21 & ALONG & ND Square 430R & MASK430R & F335M \\
\enddata
\end{deluxetable*}

\begin{deluxetable*}{lrccccl}[h!] 
\tablewidth{0pt}
\tabletypesize{\footnotesize}
\tablecaption{Cycle 2 Observations of Hot Stars, PID 4497\label{tab:4497}}
\tablehead{\colhead{Target} & \colhead{Obs\#} & \colhead{Date} & \colhead{Detector(s)} & \colhead{Subarray} & \colhead{Pupil} & \colhead{Filters}}
\startdata
\multicolumn{7}{c}{----- Part 2 Imaging -----}\\
GD 153  & 15  & 2024 Jun 02 & B1--B4 & SUB160  & CLEAR & F070W, F090W, F115W, F150W, F200W \\
       &     &  & BLONG  & SUB160  & CLEAR & F250M, F300M, F335M, F410M, F430M\\
GD 153  & 16  & 2024 Jun 05 & B1     & SUB160P & CLEAR & F187N, F212N, F210M \\
       &     &  & BLONG  & SUB160P & CLEAR & F277W, F356W, F444W\\
GD 153  & 17  & 2024 May 13 & A1--A4 & SUB160  & CLEAR & F070W, F090W, F115W, F150W, F200W \\
       &     &  & ALONG  & SUB160  & CLEAR & F250M, F300M, F335M, F410M, F430M\\
GD 153  & 18  & 2024 June 23 & A3     & SUB160P & CLEAR & F187N, F212N, F210M \\
       &     &  & ALONG  & SUB160P & CLEAR & F277W, F356W, F444W\\
LDS 749B  & 11  & 2023 Oct 13 & A3     & SUB160P & CLEAR & F070W, F090W, F115W, F150W, F200W \\
         &     &  & ALONG  & SUB160P & CLEAR & F277W, F356W, F444W, F250M, F300M \\
WD1057$+$719 & 12  & 2024 Feb 23 & A3     & SUB160P & CLEAR & F070W, F090W, F115W, F150W, F200W \\
             &     &  & ALONG  & SUB160P & CLEAR & F277W, F356W, F444W, F250M, F300M \\
\hline
\multicolumn{7}{c}{----- Part 1 Coronagraphic TA Imaging -----}\\
10 Lac  & 19 & 2023 Oct 11 & A4 & ND Square SWB & MASKSWB & F210M \\
10 Lac  & 20 & 2023 Nov 02 & A4 & ND Square SWBS & MASKSWB & F210M \\
10 Lac  & 21 & 2023 Nov 06 & A2 & ND Square 210R & MASK210R & F210M \\
10 Lac  & 22 & 2023 Nov 10 & ALONG & ND Square LWB & MASKLWB & F335M \\
10 Lac  & 23 & 2023 Nov 22 & ALONG & ND Square LWBL & MASKLWB & F335M \\
10 Lac  & 24 & 2023 Dec 20 & ALONG & ND Square 335R & MASK335R & F335M \\
10 Lac  & 25 & 2023 Dec 20 & ALONG & ND Square 430R & MASK430R & F335M \\
\enddata
\end{deluxetable*}

\begin{deluxetable*}{lrccccl}[h!] 
\tablewidth{0pt}
\tabletypesize{\footnotesize}
\tablecaption{Cycle 2 Observations of Solar Analog Stars, PID 4498\label{tab:4498}}
\tablehead{\colhead{Target} & \colhead{Obs\#} & \colhead{Date} & \colhead{Detector(s)} & \colhead{Subarray} & \colhead{Pupil\tablenotemark{a}} & \colhead{Filters}}
\startdata
\multicolumn{7}{c}{----- Part 1 Imaging -----}\\
P330-E  & 34  & 2024 Jun 10 & B1--B4 & SUB160  & CLEAR & All SW except F150W2\\
             &     &  & BLONG  & SUB160  & CLEAR & All LW except F322W2, F323N, F405N, F466N\\
P330-E  & 35  & 2024 Apr 26 & B1     & SUB160P & CLEAR & F070W, F200W, all M and N filters \\
             &     &  & BLONG  & SUB160P & CLEAR & All LW except F322W2\\
P330-E  & 36  & 2024 May 05 & A1--A4 & SUB160  & CLEAR & All SW except F150W2\\
             &     &  & ALONG  & SUB160  & CLEAR & All LW except F322W2, F323N, F405N, F466N\\
P330-E  & 37  & 2024 May 01 & A3     & SUB160P & CLEAR & F070W, F200W, all M and N filters \\
             &     &  & ALONG  & SUB160P & CLEAR & All LW except F322W2\\
P330-E  & 38  & 2024 May 05 & B1     & SUB64P & CLEAR & F070W, F090W, F115W, F150W, F200W, F150W2 \\
             &     &  & BLONG  & SUB64P & CLEAR & F277W, F356W, F444W, F250M, F460M, F322W2\\
P330-E  & 39  & 2024 May 07 & A3     & SUB64P & CLEAR & F070W, F090W, F115W, F150W, F200W, F150W2 \\
             &     &  & ALONG  & SUB64P & CLEAR & F277W, F356W, F444W, F250M, F460M, F322W2\\
\hline
\multicolumn{7}{c}{----- Part 2 Imaging -----}\\
SNAP-2  & 29 & 2024 Mar 31 & A3     & SUB160P & CLEAR & F070W, F090W, F115W, F150W, F200W \\
        &    &  & ALONG  & SUB160P & CLEAR & F277W, F356W, F444W, F250M, F300M \\
C26202  & 30 & 2023 Nov 30 & A3     & SUB160P & CLEAR & F070W, F090W, F115W, F150W, F200W \\
        &    &  & ALONG  & SUB160P & CLEAR & F277W, F356W, F444W, F335M, F360M \\
\hline
\multicolumn{7}{c}{----- Part 1 Weak Lens Imaging -----}\\
P330-E  & 41 & 2024 May 07 & B1 & SUB400P & WLP8 & F150W, F200W, F140M, F182M, F210M, F187N, F212N \\
P330-E  & 60 & 2024 May 13 & A3 & FULL & CLEAR & WLP4+F212N2\\
             &    &  & A3 & FULL & WLP8  & F070W, F140M, F182M, F210M, 187N, F212N \\
P330-E  & 61 & 2024 May 13 & A1 & FULL & CLEAR & WLP4+F212N2\\
             &    &  & A1 & FULL & WLP8  & F070W, F140M, F182M, F210M, 187N, F212N \\
\hline
\multicolumn{7}{c}{----- Part 1 Coronagraphic Imaging -----}\\
P330-E  & 46 & 2024 May 13 & A4    & SUB640 & MASKSWB & F200W, F182M, F210M, F187N, F212N \\
             &    & & ALONG & SUB320 & MASKSWB & F356W, F250M, F300M, F335M, F360M \\
P330-E  & 47 & 2024 May 08 & A2    & SUB640 & MASK210R & F200W, F182M, F210M, F187N, F212N \\
             &    & & ALONG & SUB320 & MASK210R & F444W, F360M, F410M, F430M, F460M \\
P330-E  & 48 & 2024 May 13 & ALONG & SUB400X256 & MASKLWB & All LW W and M filters \\
             &    & & A4    & SUB640     & MASKLWB & F200W, F182M, F210M, F187N, F212N \\
P330-E  & 49 & 2024 May 08 & ALONG & SUB320 & MASK335R & All LW W2, W, and M filters except F277W \\
             &    & & A2    & SUB640 & MASK335R & F200W, F182M, F210M, F187N, F212N \\
P330-E  & 50 & 2024 May 09 & ALONG & SUB320 & MASK430R & All LW W2, W, and M filters except F277W \\
             &    & & A2    & SUB640 & MASK430R & F200W, F182M, F210M, F187N, F212N \\
\hline
\multicolumn{7}{c}{----- Part 1 Coronagraphic TA Imaging -----}\\
HR 6538  & 51 & 2024 Jun 16 & A4 & ND Square SWB & MASKSWB & F210M \\
HR 6538  & 52 & 2024 Jun 04 & A4 & ND Square SWBS & MASKSWB & F210M \\
HR 6538  & 53 & 2024 Jun 04 & A2 & ND Square 210R & MASK210R & F210M \\
HR 6538  & 54 & 2023 Aug 31 & ALONG & ND Square LWB & MASKLWB & F335M \\
HR 6538  & 55 & 2023 Aug 31 & ALONG & ND Square LWBL & MASKLWB & F335M \\
HR 6538  & 56 & 2023 Aug 31 & ALONG & ND Square 335R & MASK335R & F335M \\
HR 6538  & 57 & 2023 Aug 21 & ALONG & ND Square 430R & MASK430R & F335M \\
\enddata
\tablenotetext{a}{When not using WLP8 or a coronagraphic mask, most filters are paired with the CLEAR element in the pupil wheel. The exception is filters that are {\em in} the pupil wheel, which are paired with wide filters in the filter wheel (Pupil$+$Filter: F164M$+$F150W2, F162N$+$F150W2, F323N$+$F322W2, F405N$+$F444W, F460N$+$F444W, and F470N$+$F444W).}
\end{deluxetable*}

\begin{deluxetable*}{lrccccl}[h!] 
\tablewidth{0pt}
\tabletypesize{\footnotesize}
\tablecaption{Cycle 3 Observations of Hot Stars, PID 6605\label{tab:6605}}
\tablehead{\colhead{Target} & \colhead{Obs\#} & \colhead{Date} & \colhead{Detector(s)} & \colhead{Subarray} & \colhead{Pupil\tablenotemark{a}} & \colhead{Filters}}
\startdata
\multicolumn{7}{c}{----- Part 1 Imaging -----}\\
GD 71  & 1   & 2024 Oct 06 & B1--B4 & SUB160  & CLEAR & All SW filters \\
       &     &  & BLONG  & SUB160  & CLEAR & All LW except F323N, F405N, F466N\\
GD 71  & 2   & 2024 Oct 09 & B1     & SUB160P & CLEAR & All SW except F150W2 \\
       &     &  & BLONG  & SUB160P & CLEAR & All LW filters \\
GD 71  & 3   & 2024 Nov 01 & A1--A4 & SUB160  & CLEAR & All SW filters \\
       &     &  & ALONG  & SUB160  & CLEAR & All LW except F323N, F405N, F466N\\
GD 71  & 4   &  2024 Oct 22 & A3     & SUB160P & CLEAR & All SW except F150W2 \\
       &     &  & ALONG  & SUB160P & CLEAR & All LW filters\\
\hline
\multicolumn{7}{c}{----- Faint White Dwarfs -----}\\
WDFS0122$-$30 & 105  & 2024 Dec 17 & B1     & FULL    & CLEAR & F070W, F090W, F115W, F150W, F200W \\
              &     &  & BLONG  & FULL    & CLEAR & F277W, F356W, F444W, F250M, F480M \\
WDFS0458$-$56 & 6  & 2024 Aug 14 & B1     & FULL & CLEAR & F070W, F090W, F115W, F150W, F200W \\
             &     &  & BLONG  & FULL & CLEAR & F277W, F356W, F444W, F250M, F480M \\
WDFS2317$-$29 & 9  & 2024 Oct 12 & A3     & FULL & CLEAR & F070W, F090W, F115W, F150W, F200W \\
             &     &  & ALONG  & FULL & CLEAR & F277W, F356W, F444W, F250M, F480M \\
\enddata
\tablenotetext{a}{When not using WLP8 or a coronagraphic mask, most filters are paired with the CLEAR element in the pupil wheel. The exception is filters that are {\em in} the pupil wheel, which are paired with wide filters in the filter wheel (Pupil$+$Filter: F164M$+$F150W2, F162N$+$F150W2, F323N$+$F322W2, F405N$+$F444W, F460N$+$F444W, and F470N$+$F444W).}
\end{deluxetable*}

\begin{deluxetable*}{lrccccl}[h!] 
\tablewidth{0pt}
\tabletypesize{\footnotesize}
\tablecaption{Cycle 3 Observations of Solar Analog Stars, PID 6606\label{tab:6606}}
\tablehead{\colhead{Target} & \colhead{Obs\#} & \colhead{Date} & \colhead{Detector(s)} & \colhead{Subarray} & \colhead{Pupil} & \colhead{Filters}}
\startdata
\multicolumn{7}{c}{----- Part 1 Weak Lens Imaging -----}\\
P330-E  & 1  & 2025 Mar 03 & B1    & SUB400P & WLP8 & F150W, F200W, F140M, F182M, F210M, F187N, F212N \\
P330-E  & 4  & 2025 Mar 03 & A3 & FULL & CLEAR & WLP4+F212N2\\
             &    &  & A3 & FULL & WLP8  & F070W, F140M, F182M, F210M, 187N, F212N \\
P330-E  & 5  & 2024 Aug 22 & A1 & FULL & CLEAR & WLP4+F212N2\\
             &    &  & A1 & FULL & WLP8  & F070W, F140M, F182M, F210M, 187N, F212N \\
P330-E  & 13  & 2025 Mar 03 & A1 & SUBGRISM256 & CLEAR & WLP4+F212N2\\
P330-E  & 14  & 2025 Mar 03 & A3 & SUBGRISM256 & CLEAR & WLP4+F212N2\\
\hline
\multicolumn{7}{c}{----- Part 1 Coronagraphic Imaging -----}\\
P330-E  & 8  & 2024 Aug 03 & A4    & SUB640 & MASKSWB & F200W, F182M, F210M, F187N, F212N \\
             &    & & ALONG & SUB320 & MASKSWB & F356W, F250M, F300M, F335M, F360M \\
P330-E  & 9  & 2025 Mar 04 & A2    & SUB640 & MASK210R & F200W, F182M, F210M, F187N, F212N \\
             &    & & ALONG & SUB320 & MASK210R & F444W, F360M, F410M, F430M, F460M \\
P330-E  & 10 & 2025 May 10& ALONG & SUB400X256 & MASKLWB & All LW W and M filters \\
             &    & & A4    & SUB640     & MASKLWB & F200W, F182M, F210M, F187N, F212N \\
P330-E  & 11 & 2025 Mar 05& ALONG & SUB320 & MASK335R & All LW W2, W, and M filters except F277W \\
             &    & & A2    & SUB640 & MASK335R & F200W, F182M, F210M, F187N, F212N \\
P330-E  & 12 & 2024 Aug 05& ALONG & SUB320 & MASK430R & All LW W2, W, and M filters except F277W \\
             &    & & A2    & SUB640 & MASK430R & F200W, F182M, F210M, F187N, F212N \\
\hline
\multicolumn{7}{c}{----- Part 1 Coronagraphic TA Imaging -----}\\
HR 6538  & 15 & 2025 Mar 27 & A4 & ND Square SWB & MASKSWB & F210M \\
HR 6538  & 16 & 2025 Mar 26 & A4 & ND Square SWBS & MASKSWB & F210M \\
HR 6538  & 17 & 2025 Mar 26 & A2 & ND Square 210R & MASK210R & F210M \\
HR 6538  & 18 & 2024 Aug 18 & ALONG & ND Square LWB & MASKLWB & F335M \\
HR 6538  & 19 & 2025 Mar 26 & ALONG & ND Square LWBL & MASKLWB & F335M \\
HR 6538  & 20 & 2024 Jul 29 & ALONG & ND Square 335R & MASK335R & F335M \\
HR 6538  & 21 & 2024 Sep 07 & ALONG & ND Square 430R & MASK430R & F335M \\
\enddata
\end{deluxetable*}

\begin{deluxetable*}{lrccccl}[h!] 
\tablewidth{0pt}
\tabletypesize{\footnotesize}
\tablecaption{Cycle 4 Observations of Solar Analog Stars, PID 7615\label{tab:7615}}
\tablehead{\colhead{Target} & \colhead{Obs\#} & \colhead{Date} & \colhead{Detector(s)} & \colhead{Subarray} & \colhead{Pupil\tablenotemark{a}} & \colhead{Filters}}
\startdata
\multicolumn{7}{c}{----- Part 1 Weak Lens Imaging -----}\\
P330-E  & 1  & 2025 Jul 08 & B1    & SUB400P & WLP8 & F150W, F200W, F140M, F182M, F210M, F187N, F212N \\
P330-E  & 4  & 2025 Jul 21 & A3 & FULL & CLEAR & WLP4+F212N2\\
             &    &  & A3 & FULL & WLP8  & F200W, F140M, F182M, F210M, 187N, F212N \\
P330-E  & 5  & 2025 Aug 04 & A1 & FULL & CLEAR & WLP4+F212N2\\
             &    &  & A1 & FULL & WLP8  & F200W, F140M, F182M, F210M, 187N, F212N \\
P330-E  & 8  & 2025 Aug 23 & A1 & SUBGRISM256 & CLEAR & WLP4+F212N2\\
P330-E  & 9  & 2025 Jul 08 & A3 & SUBGRISM256 & CLEAR & WLP4+F212N2\\
\hline
\multicolumn{7}{c}{----- Part 1 Coronagraphic Imaging -----}\\
P330-E  & 10  & 2025 Jul 21 & A2    & FULL & MASK210R & F200W, F182M, F210M, F187N, F212N \\
             &    & & ALONG & FULL & MASK210R & F356W, F444W, F250M, F300M, F335M, \\
             &    & &       &      &          & F360M, F410M, F430M, F460M, F480M, F322W2 \\
P330-E  & 11  & 2025 Jul 11 & A4    & FULL & MASKSWB & F200W, F182M, F210M, F187N, F212N \\
             &    & & ALONG & FULL & MASKSWB & F277W, F356W, F444W, F250M, F300M, \\
             &    & &       &      &         & F335M, F360M, F410M, F430M, F460M, F480M \\
\hline
\multicolumn{7}{c}{----- Part 1 Coronagraphic TA Imaging -----}\\
HR 6538  & 12 & 2025 Jul 8 & A4 & ND Square SWB & MASKSWB & F210M \\
HR 6538  & 13 & 2025 Jul 8 & A4 & ND Square SWBS & MASKSWB & F210M \\
HR 6538  & 14 & 2025 Jul 8 & A2 & ND Square 210R & MASK210R & F210M \\
HR 6538  & 15 & 2025 Jul 8 & ALONG & ND Square LWB & MASKLWB & F335M \\
HR 6538  & 16 & 2025 Jul 8 & ALONG & ND Square LWBL & MASKLWB & F335M \\
HR 6538  & 17 & 2025 Jul 8 & ALONG & ND Square 335R & MASK335R & F335M \\
HR 6538  & 18 & 2025 Jul 8 & ALONG & ND Square 430R & MASK430R & F335M \\
\enddata
\end{deluxetable*}

\begin{deluxetable*}{lrccccl}[h!] 
\tablewidth{0pt}
\tabletypesize{\footnotesize}
\tablecaption{Cycle 4 Observations of A dwarf Stars, PID 7487\label{tab:7487}}
\tablehead{\colhead{Target} & \colhead{Obs\#} & \colhead{Date} & \colhead{Detector(s)} & \colhead{Subarray} & \colhead{Pupil\tablenotemark{a}} & \colhead{Filters}}
\startdata
\multicolumn{7}{c}{----- Part 1 Coronagraphic Imaging -----}\\
J1743045  & 1  & 2025 Jul 04 & A2    & FULL & MASK210R & F200W, F182M, F210M, F187N, F212N \\
             &    & & ALONG & FULL & MASK210R & F356W, F444W, F250M, F300M, F335M, \\
             &    & &       &        &          & F360M, F410M, F430M, F460M, F480M, F322W2 \\
J1743045  & 2  & 2025 Sep 17 & A4    & FULL & MASKSWB & F200W, F182M, F210M, F187N, F212N \\
             &    & & ALONG & FULL & MASKSWB & F277W, F356W, F444W, F250M, F300M, \\
             &    & &       &        &          & F335M, F360M, F410M, F430M, F460M, F480M \\
\enddata
\end{deluxetable*}

\begin{deluxetable*}{llcl}[tbp]
\tablecaption{Repeatability Observations\label{tab:1539}}
\tablehead{\colhead{Obs\#} & \colhead{Date} & \colhead{Detector(s)} & \colhead{Subarray}}
\startdata
\multicolumn{4}{c}{----- PID 1539 -----}\\
8, 9 &2022 Jul 19 & All & SUB160 \\
74, 75 &2022 Aug 30 & All & SUB160 \\
24, 25 &2022 Sep 16 & All & SUB160 \\
29, 30 &2023 Jan 05 & All & SUB160 \\
34, 35 &2023 Feb 09 & All & SUB160 \\
39, 40 &2023 Mar 10 & All & SUB160 \\
44, 45 &2023 Apr 06 & All & SUB160 \\
49, 50 &2023 May 03 & All & SUB160 \\
54, 55 &2023 May 30 & All & SUB160 \\
59, 60 &2023 Jun 30 & All & SUB160 \\
\hline
\multicolumn{4}{c}{----- PID 4499 -----}\\
7, 8 &2023 Jul 06 & All SW & SUB160 \\
6, 9 &2023 Jul 06 & All LW & SUB160P \\
14, 15 &2023 Jul 30 & All SW & SUB160 \\
13, 16 &2023 Jul 30 & All LW & SUB160P \\
21, 22 &2023 Sep 05 & All SW & SUB160 \\
20, 23 &2023 Sep 05 & All LW & SUB160P \\
28, 29 &2023 Sep 26 & All SW & SUB160 \\
27, 30 &2023 Sep 26 & All LW & SUB160P\\
35, 36 &2024 Jan 10 & All SW & SUB160 \\
34, 37 &2024 Jan 10 & All LW & SUB160P\\
42, 43 &2024 Feb 07 & All SW & SUB160 \\
41, 44 &2024 Feb 07 & All LW & SUB160P\\
49, 50 &2024 Mar 03 & All SW & SUB160 \\
48, 51 &2024 Mar 03 & All LW & SUB160P\\
56, 57 &2024 Mar 31 & All SW & SUB160 \\
55, 58 &2024 Mar 31 & All LW & SUB160P\\
63, 64 &2024 Apr 28 & All SW & SUB160 \\
62, 65 &2024 Apr 28 & All LW & SUB160P\\
70, 71 &2024 May 26 & All SW & SUB160 \\
69, 72 &2024 May 26 & All LW & SUB160P\\
77, 78 &2024 Jun 23 & All SW & SUB160 \\
76, 79 &2024 Jun 23 & All LW & SUB160P\\
\hline
\multicolumn{4}{c}{----- PID 6607 -----}\\
1, 2 & 2024 Jul 26 & All LW & SUB160P \\
4\tablenotemark{a} & 2024 Jul 26 & B1--B4 & SUB160 \\
19, 20 & 2024 Sep 25 & All LW & SUB160P \\
22\tablenotemark{a} & 2024 Sep 25 & B1-B4 & SUB160 \\
35, 36 & 2025 Feb 05 & All LW & SUB160P \\
37, 38 & 2025 Feb 05 & All SW & SUB160 \\
51, 52 & 2025 Apr 01 & All LW & SUB160P \\
53, 54 & 2025 Apr 01 & All SW & SUB160 \\
67, 68 & 2025 May 28 & All LW & SUB160P \\
69, 70 & 2025 May 28 & All SW & SUB160 \\
\hline
\multicolumn{4}{c}{----- PID 7671 -----}\\
1, 2 & 2025 Jul 24 & All LW & SUB160P \\
3, 4 & 2025 Jul 24 & All SW & SUB160 \\
12, 13 & 2025 Sep 24 & All LW & SUB160P \\
14, 15 & 2025 Sep 24 & All SW & SUB160
\enddata
\tablecomments{The target for all observations is BD$+$60\,1753. All SW observations use the F212N filter and all LW observations use the F470N filter. }
\tablenotetext{a}{Obs 3 and 21 missed the target on A1--A4.}
\end{deluxetable*}

\begin{deluxetable*}{lllll}[tbp]
\tablewidth{0pt}
\tabletypesize{\footnotesize}
\tablecaption{FULL-to-Subarray Flux Offsets, PID 4452 \label{tab:4452}}
\tablehead{\colhead{Target} & \colhead{Obs\#} & \colhead{Date} & \colhead{Detector(s)} & \colhead{Subarray}}
\startdata
C26206 & 1--4 & 2023 Aug 10 & B1, BLONG & SUB64P, SUB160P, SUB400P, FULL \\
C26206 & 5--8 &2023 Aug 20 & BLONG & SUB160, SUB320, SUB640, FULL \\
C26206 & 9--12 &2023 Jul 31 & B1 & SUB160, SUB320, SUB640, FULL \\
C26206 & 13--16 &2023 Jul 31 & B2 & SUB160, SUB320, SUB640, FULL \\
C26206 & 17--20 &2023 Jul 31 & B3 & SUB160, SUB320, SUB640, FULL \\
C26206 & 21--24 &2023 Jul 31 & B4 & SUB160, SUB320, SUB640, FULL \\
C26206 & 25--26 &2023 Jul 31 & A3, ALONG & SUB64P, FULL \\
C26206 & 27--28 &2023 Jul 31 & ALONG & SUB160, FULL \\
C26206 & 29--30 &2023 Jul 31 & A1 & SUB160, FULL \\
C26206 & 31--32 &2023 Jul 31 & A2 & SUB160, FULL \\
C26206 & 33--34 &2023 Jul 31 & A3 & SUB160, FULL \\
C26206 & 35--36 &2023 Jul 31 & A4 & SUB160, FULL \\
C26206 & 37--40 &2023 Jul 31 & A1, ALONG & SUBGRISM64, SUBGRISM128, SUBGRISM256, FULL \\
C26206 & 41--44 &2023 Jul 31 & A3, ALONG & SUBGRISM64, SUBGRISM128, SUBGRISM256, FULL \\
C26206 & 45--46 &2023 Jul 31 & ALONG & SUBGRISM64, FULL \\
C26206 & 47--48 &2023 Dec 30 & A2, ALONG & SUB640, FULL (MASK210R) \\
C26206 & 49--50 &2023 Dec 15 & A4, ALONG & SUB640, FULL (MASKSWB) \\
C26206 & 51--52 &2023 Dec 29 & A2, ALONG & SUB320, FULL (MASK335R) \\
C26206 & 53--54 &2023 Dec 30 & A2, ALONG & SUB320, FULL (MASK430R) \\
C26206 & 55--56 &2023 Jul 31 & A4, ALONG & SUB400x256, FULL (MASKLWB) \\
\enddata
\tablecomments{SW observations use filter F210M, LW observations use filter F410M. For coronagraphy, the LW filter is F356W.}
\end{deluxetable*}

\begin{deluxetable*}{lllll}[tbp]
\tablewidth{0pt}
\tabletypesize{\footnotesize}
\tablecaption{FULL-to-Subarray Flux Offsets, PID 6630 \label{tab:6630}}
\tablehead{\colhead{Target} & \colhead{Obs\#} & \colhead{Date} & \colhead{Detector(s)} & \colhead{Subarray}}
\startdata
C26206 & 1--4 & 2024 Oct 22 & B1, BLONG & SUB64P, SUB160P, SUB400P, FULL \\
C26206 & 5--8 &2025 Jan 18 & BLONG & SUB160, SUB320, SUB640, FULL \\
C26206 & 9--12 &2025 Jan 07 & B1 & SUB160, SUB320, SUB640, FULL \\
C26206 & 13--16 &2024 Nov 14 & B2 & SUB160, SUB320, SUB640, FULL \\
C26206 & 17--20 &2025 Jan 07 & B3 & SUB160, SUB320, SUB640, FULL \\
C26206 & 21--24 &2024 Oct 22 & B4 & SUB160, SUB320, SUB640, FULL \\
C26206 & 25--26 &2024 Oct 22 & A3, ALONG & SUB64P, FULL \\
C26206 & 27--28 &2025 Jan 15 & ALONG & SUB160, FULL \\
C26206 & 31--32 &2025 Jan 07 & A2 & SUB160, FULL \\
C26206 & 33--34 &2024 Oct 22 & A3 & SUB160, FULL \\
C26206 & 35--36 &2025 Jan 07 & A4 & SUB160, FULL \\
C26206 & 37--40 &2024 Nov 27 & A1, ALONG & SUBGRISM64, SUBGRISM128, SUBGRISM256, FULL \\
C26206 & 41--44 &2025 Jan 07 & A3, ALONG & SUBGRISM64, SUBGRISM128, SUBGRISM256, FULL \\
C26206 & 45--46 &2025 Jan 07 & ALONG & SUBGRISM64, FULL \\
P330-E & 47--48 &2025 Feb 27 & A2, ALONG & SUB640, FULL (MASK210R) \\
P330-E & 49--50 &2025 Mar 01 & A4, ALONG & SUB640, FULL (MASKSWB) \\
P330-E & 51--52 &2025 Mar 01 & A2, ALONG & SUB320, FULL (MASK335R) \\
P330-E & 53--54 &2025 Mar 01 & A2, ALONG & SUB320, FULL (MASK430R) \\
P330-E & 55--56 &2025 Mar 02 & A4, ALONG & SUB400x256, FULL (MASKLWB) \\
\enddata
\tablecomments{SW observations use filter F210M, LW observations use filter F410M. For coronagraphy, we use SW filter F200W and LW filter F356W. }
\end{deluxetable*}

\begin{deluxetable*}{lllll}[tbp]
\tablewidth{0pt}
\tabletypesize{\footnotesize}
\tablecaption{FULL-to-Subarray Flux Offsets, PID 8882 \label{tab:8882}}
\tablehead{\colhead{Target} & \colhead{Obs\#} & \colhead{Date} & \colhead{Detector(s)} & \colhead{Subarray}}
\startdata
C26206 & 1--4 & 2025 Aug 19 & B1, BLONG & SUB64P, SUB160P, SUB400P, FULL \\
C26206 & 5--8 &2025 Aug 26 & BLONG & SUB160, SUB320, SUB640, FULL \\
C26206 & 9, 11, 12 &2025 Sep 22 & B1 & SUB160, SUB640, FULL \\
C26206 & 13, 15, 16 &2025 Sep 22 & B2 & SUB160, SUB640, FULL \\
C26206 & 17, 19, 20 &2025 Oct 12 & B3 & SUB160, SUB640, FULL \\
C26206 & 21, 23, 24 &2025 Oct 15 & B4 & SUB160, SUB640, FULL \\
C26206 & 157, 158 &2025 Nov 29 & A3, ALONG & SUB160P, FULL\\
C26206 & 27, 28, 55 &2025 Oct 24 & ALONG & SUB160, SUB320, FULL \\
C26206 & 29, 30, 54 &2025 Nov 20 & A1 & SUB160, SUB320, FULL \\
C26206 & 31--32 &2025 Nov 29 & A2 & SUB160, FULL \\
C26206 & 33, 34, 56 &2025 Dec 10 & A3 & SUB160, SUB320, FULL \\
C26206 & 35--36 &2025 Dec 10 & A4 & SUB160, FULL \\
C26206 & 37, 38, 40 &2025 Dec 09 & A1, ALONG & SUBGRISM64, SUBGRISM128, FULL \\
P330-E & 47--48 &2025 Aug 10 & A2, ALONG & SUB640, FULL (MASK210R) \\
P330-E & 51--52 &2025 Aug 14 & A2, ALONG & SUB320, FULL (MASK335R) \\
\enddata
\tablecomments{Observations for this program had not fully executed at the time of this analysis; we list only observations included here.  SW observations use filter F210M, LW observations use filter F410M. For coronagraphy, we use SW filter F200W and LW filter F356W. }
\end{deluxetable*}

\end{document}